\documentclass[sigconf]{acmart}

\AtBeginDocument{%
  }

\newif\ifPreprint
\Preprinttrue

\ifPreprint
  \setcopyright{none}
  \copyrightyear{2026}
  \acmYear{2026}
  \renewcommand\footnotetextcopyrightpermission[1]{}
  \acmDOI{}
  \acmISBN{}
\else
  \setcopyright{acmlicensed}
  \copyrightyear{2027}
  \acmYear{2027}
  \acmDOI{XXXXXXX.XXXXXXX}
  \acmConference[CHI '27]{Proceedings of the 2027 CHI Conference on Human Factors in Computing Systems}{April 26--May 01, 2027}{Yokohama, Japan}
  \acmISBN{978-1-4503-XXXX-X/2027/04}
\fi

\newcommand{\nReviews}{1,577}
\newcommand{\nBooks}{78}
\newcommand{\nCategories}{8}
\newcommand{\nLowStar}{863}

\newcommand{\oneStarShare}{51\%}
\newcommand{\nNonUsReviews}{77}
\newcommand{\nonUsPct}{5.1\%}

\newcommand{\nFlagged}{38}
\newcommand{\nNotFlagged}{825}
\newcommand{\overallFlagRate}{4.4\%}
\newcommand{\nHeuristicExcluded}{2}
\newcommand{\nHeuristicCoded}{36}
\newcommand{\heuristicKappa}{0.801}
\newcommand{\heuristicExactN}{30}
\newcommand{\heuristicExactPct}{83\%}
\newcommand{\imSamplePct}{20\%}
\newcommand{\imNSampled}{379}
\newcommand{\imNPairs}{1{,}753}
\newcommand{\imKappaOverall}{0.804}
\newcommand{\imPearsonOverall}{0.914}
\newcommand{\imWithinOne}{76.4\%}
\newcommand{\imKappaAI}{0.939}
\newcommand{\imExactAI}{99.2\%}

\newcommand{\imBias}{0.28}
\newcommand{\imIccMin}{0.683}
\newcommand{\imLowStarIccMin}{0.846}
\newcommand{\imIccGenAI}{0.912}
\newcommand{\imWithinOneGenAI}{94.1\%}
\newcommand{\heuristicWeightedKappa}{0.819}
\newcommand{\dZeroSharePct}{19.4\%}
\newcommand{\dZeroCount}{7}
\newcommand{\dOneCount}{8}
\newcommand{\dTwoCount}{7}
\newcommand{\dOneTwoCooccur}{8}
\newcommand{\crosstabCQrrb}{0.49}
\newcommand{\crosstabPRESrrb}{0.41}
\newcommand{\crosstabMETArrb}{0.36}
\newcommand{\exGenAiCQd}{0.83}
\newcommand{\exGenAiPRESd}{0.93}
\newcommand{\exGenAiMETAd}{-0.70}
\newcommand{\exGenAiFlagsRemoved}{13}
\newcommand{\mostFlaggedFlagged}{9}
\newcommand{\mostFlaggedLowStar}{18}
\newcommand{\nextFlaggedPct}{27\%}
\newcommand{\nextFlaggedFlagged}{3}
\newcommand{\nextFlaggedLowStar}{11}

\newcommand{\mostFlaggedReviews}{4{,}603}
\newcommand{\mostFlaggedRating}{4.1}
\newcommand{\mostFlaggedRank}{2}
\newcommand{\mostFlaggedFlagPct}{50\%}

\newcommand{\genAiFlagRate}{35.1\%}

\newcommand{\hOneH}{95.66}
\newcommand{\hOneDf}{7}
\newcommand{\hOneEpsSq}{0.111}
\newcommand{\hOneEffect}{medium}

\newcommand{\hEightChi}{93.58}
\newcommand{\hEightDf}{7}
\newcommand{\hEightV}{0.329}
\newcommand{\hEightEffect}{medium}

\newcommand{\hTwoU}{83,308}
\newcommand{\hTwoP}{0.002}
\newcommand{\hTwoD}{0.21}
\newcommand{\hTwoAiN}{267}
\newcommand{\hTwoAiM}{1.61}
\newcommand{\hTwoNonAiN}{596}
\newcommand{\hTwoNonAiM}{1.25}

\newcommand{\hFourD}{1.10}
\newcommand{\hFourP}{0.002}
\newcommand{\hFourU}{557.5}
\newcommand{\hFourIndieN}{15}
\newcommand{\hFourIndieM}{1.80}
\newcommand{\hFourEstabN}{52}
\newcommand{\hFourEstabM}{1.15}
\newcommand{\hFiveRho}{0.155}
\newcommand{\hFiveP}{0.198}
\newcommand{\hFiveN}{71}
\newcommand{\hSevenRho}{0.086}

\newcommand{\hNineFlagN}{20}
\newcommand{\hNineCleanN}{51}
\newcommand{\hNineFlagRank}{5.5}
\newcommand{\hNineCleanRank}{7.0}
\newcommand{\hNineFlagRankPct}{0.47}
\newcommand{\hNineCleanRankPct}{0.60}
\newcommand{\hNineU}{469.0}
\newcommand{\hNineP}{0.604}
\newcommand{\hNineUPct}{451.0}
\newcommand{\hNinePPct}{0.454}
\newcommand{\hNineTopThreeFlagged}{7}
\newcommand{\hNineTopThree}{20}
\newcommand{\hSevenP}{0.478}
\newcommand{\hSevenN}{71}

\newcommand{\crosstabCQd}{0.90}
\newcommand{\crosstabPRESd}{0.78}
\newcommand{\crosstabMETAd}{-0.66}

\newcommand{\iccOverall}{0.902}
\newcommand{\iccAI}{0.969}
\newcommand{\dimMeanMetaIssues}{5.56}
\newcommand{\dimMeanContentQuality}{4.35}
\newcommand{\dimMeanPresentation}{3.06}
\newcommand{\dimMeanAccuracy}{2.02}
\newcommand{\dimMeanAiPerception}{1.36}
\newcommand{\genAiMean}{3.81}

\newcommand{\hThreeContentH}{48.79}
\newcommand{\hThreeContentDf}{7}
\newcommand{\hThreeContentEps}{0.057}
\newcommand{\hThreeContentEffect}{small}

\newcommand{\flaggedOneStarPct}{68.4\%}
\newcommand{\nonflaggedOneStarPct}{50.1\%}

\newcommand{\nFlaggedBooks}{20}

\newcommand{\flaggedIndiePct}{42\%}
\newcommand{\flaggedIndieN}{19}

\newcommand{\nTwoFlagBooks}{8}
\newcommand{\nProdMultiBooks}{4}
\newcommand{\nNoLowStarBooks}{7}
\newcommand{\nNoLowStarClassified}{5}
\newcommand{\nPublisherExcluded}{11}
\newcommand{\nPublishers}{60}
\newcommand{\publisherAgreement}{96.7\%}
\newcommand{\nPublisherAdjudicated}{12}
\newcommand{\nPublisherUnresolved}{6}

\newcommand{\nMultiFlagBooks}{3}
\newcommand{\multiFlagIndiePct}{67\%}
\newcommand{\nEstabFlagged}{11}
\newcommand{\estabMaxFlags}{4}
\newcommand{\flagRateAIAndSemantics}{3.1\%}
\newcommand{\flagRateMachineTheory}{3.0\%}

\newcommand{\flagRateComputerSecurity}{5.4\%}
\newcommand{\flagRateProgLanguages}{1.0\%}
\newcommand{\flagRateGardening}{5.7\%}
\newcommand{\genAiToNextRatio}{6.2$\times$}
\newcommand{\nextHighestCat}{Gardening}
\newcommand{\nextHighestRate}{5.7\%}

\newcommand{\proxMedN}{215}

\newcommand{\proxHighMean}{1.61}
\newcommand{\proxMedMean}{1.27}
\newcommand{\proxZeroMean}{1.24}
\newcommand{\proxKwH}{8.87}
\newcommand{\proxKwP}{0.012}
\newcommand{\proxRho}{0.095}
\newcommand{\proxRhoP}{0.005}
\newcommand{\proxHzU}{53,426}
\newcommand{\proxHzP}{0.005}
\newcommand{\proxHzD}{0.20}
\newcommand{\proxHmU}{29,882}
\newcommand{\proxHmP}{0.068}
\newcommand{\proxHmD}{0.19}
\newcommand{\proxMzU}{41,364}
\newcommand{\proxMzP}{0.529}
\newcommand{\proxMzD}{0.01}

\newcommand{\recHighBookPct}{82\%}
\newcommand{\recMedBookPct}{55\%}
\newcommand{\recZeroBookPct}{93\%}
\newcommand{\recHighN}{194}
\newcommand{\recMedN}{97}
\newcommand{\recZeroN}{349}
\newcommand{\recHighFlagRate}{9.3\%}
\newcommand{\recMedFlagRate}{7.2\%}
\newcommand{\recZeroFlagRate}{2.9\%}
\newcommand{\recMzU}{17,944}
\newcommand{\recMzP}{0.012}
\newcommand{\recMzD}{0.19}

\newcommand{\recHmP}{0.761}

\newcommand{\preChatGptBookFlags}{2}
\newcommand{\preChatGptReviewFlags}{1}
\newcommand{\nFlaggedFalsePos}{2}
\newcommand{\flaggedFalsePosPct}{5.3\%}

\newcommand{\shallowFlagHigh}{13.0\%}
\newcommand{\shallowNHigh}{123}
\newcommand{\shallowFlagMed}{7.0\%}
\newcommand{\shallowNMed}{57}
\newcommand{\shallowFlagZero}{4.4\%}
\newcommand{\shallowNZero}{136}
\newcommand{\shallowChi}{6.46}
\newcommand{\shallowChiDf}{2}
\newcommand{\shallowChiP}{0.040}
\newcommand{\shallowOR}{3.24}
\newcommand{\shallowORP}{0.015}

\newcommand{\humanValN}{50}
\newcommand{\humanValFlaggedN}{38}
\newcommand{\humanValNonFlaggedN}{12}
\newcommand{\humanIccCQ}{0.799}
\newcommand{\humanIccCQciLo}{0.67}
\newcommand{\humanIccCQciHi}{0.88}
\newcommand{\humanIccAcc}{0.621}
\newcommand{\humanIccAccciLo}{0.39}
\newcommand{\humanIccAccciHi}{0.77}
\newcommand{\humanIccAI}{0.924}
\newcommand{\humanIccAIciLo}{0.87}
\newcommand{\humanIccAIciHi}{0.96}
\newcommand{\humanIccPres}{0.526}
\newcommand{\humanIccPresciLo}{0.30}
\newcommand{\humanIccPresciHi}{0.70}
\newcommand{\humanIccMeta}{0.597}
\newcommand{\humanIccMetaciLo}{0.25}
\newcommand{\humanIccMetaciHi}{0.78}
\newcommand{\humanAiKappa}{0.947}
\newcommand{\humanAiAgreement}{98\%}
\newcommand{\humanAiTP}{37}
\newcommand{\humanAiTN}{12}
\newcommand{\humanAiFP}{1}
\newcommand{\humanAiFN}{0}

\usepackage{booktabs}
\usepackage{graphicx}
\usepackage{subcaption}
\usepackage{xcolor}
\usepackage{multirow}
\usepackage{tikz}
\usetikzlibrary{arrows.meta, positioning, fit, backgrounds, calc, shapes.misc}

\definecolor{tierHigh}{HTML}{B18DFF}       % Purple - AI/ML categories
\definecolor{tierHighLight}{HTML}{E8DCFF}  % Light purple - backgrounds
\definecolor{tierMed}{HTML}{9E9E9E}        % Gray - Tech non-AI
\definecolor{tierMedLight}{HTML}{E8E8E8}   % Light gray - backgrounds
\definecolor{tierZero}{HTML}{FADA62}       % Yellow - Non-tech
\definecolor{tierZeroLight}{HTML}{FFF3CC}  % Light yellow - backgrounds
\definecolor{accent}{HTML}{333333}         % Dark text
\definecolor{connector}{HTML}{888888}      % Arrow/line color

\graphicspath{{figures/}}

\makeatletter
\newcommand{\narrowwidth}{\if@ACM@manuscript 3.4in\else\columnwidth\fi}

\newcommand{\profilefigure}{%
  \if@ACM@manuscript
    \includegraphics[width=\textwidth]{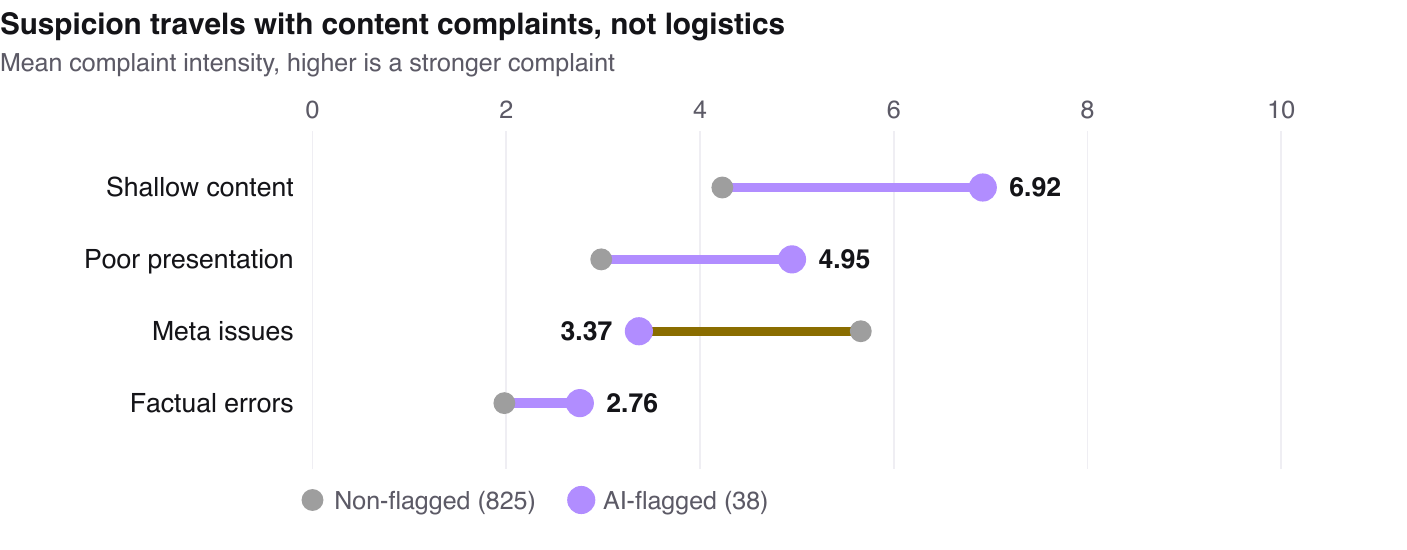}%
  \else
    \includegraphics[width=\columnwidth]{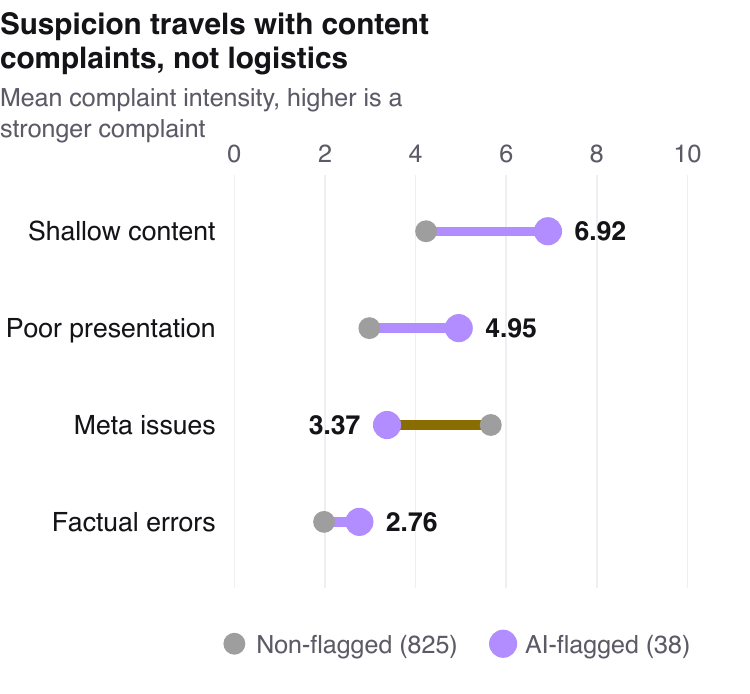}%
  \fi}
\makeatother

\begin{document}

%% Title
\title{``Is This Book AI-Generated?'' How Authorship Suspicion Manifests in Marketplace Reviews}

%% ============================================================================
%% AUTHOR BLOCK — atomic anonymous/de-anonymized switch.
%% Keep consistent with the companion data paper (datapaper/main.tex).
%%
%% ONE-CHARACTER CAMERA-READY SWAP:
%%   \Anonymoustrue   → anonymous (current; use during double-blind review)
%%   \Anonymousfalse  → de-anonymized (use for arXiv preprint + camera-ready)
%%
%% Affiliation defaults to Option 1 (MSR + Light disclaimer). To switch to
%% Option 2 (Independent / gmail), change \UseMSRtrue to \UseMSRfalse.
%% ============================================================================

\newif\ifAnonymous
\Anonymousfalse   %% ← anonymous for CHI 2027 review; flip to \Anonymousfalse for arXiv / camera-ready

\newif\ifUseMSR
\UseMSRfalse     %% ← \UseMSRtrue = Option 1 (MSR); \UseMSRfalse = Option 2 (Independent)

\ifAnonymous
  \author{Anonymous Author(s)}
  \affiliation{%
    \institution{Anonymous Institution}
    \country{}}
  \email{anon@example.com}
\else
  \ifUseMSR
    \author{Victor Dibia}
    \affiliation{%
      \institution{Microsoft Research}
      \city{Redmond}
      \state{WA}
      \country{USA}}
    \email{victordibia@microsoft.com}
    \thanks{The views expressed are those of the author and do not
      necessarily reflect the position of Microsoft Corporation.}
    \renewcommand{\shortauthors}{Dibia}
  \else
    \author{Victor Dibia}
    \affiliation{%
      \institution{Independent Researcher}
      \country{USA}}
    \email{victor.dibia@gmail.com}
    \renewcommand{\shortauthors}{Dibia}
  \fi
\fi

%% Abstract
\begin{abstract}
As AI becomes part of how books are authored, reader response to suspected AI authorship grows more consequential, yet remains unexamined. We analyze \nLowStar{} low-star reviews of \nBooks{} Amazon bestsellers across \nCategories{} categories at three levels of proximity to AI. Suspicion concentrates in Generative AI books (\genAiFlagRate{}) but appears in every category, including Gardening (\flagRateGardening{}). Reviews citing AI authorship complain more about shallow content and poor presentation than other critical reviews. Suspicion takes two forms: \textit{ambient}, where ``AI-generated'' is a generic complaint about formulaic writing, and \textit{corroborated}, where reviewers of the same book independently cite concrete evidence. We propose two mechanisms by which suspicion arises: topical concentration, where a book's AI subject matter supplies vocabulary for quality complaints, and artifact detection, where readers notice ChatGPT-style formatting regardless of topic. Star ratings can hide this suspicion: the most-flagged book holds \mostFlaggedRating{} stars while \mostFlaggedFlagPct{} of its critical reviews cite AI authorship.
\end{abstract}



\maketitle

%% Paper sections
\section{Introduction}
\label{sec:introduction}

\begin{figure*}[t]
\centering
\includegraphics[width=\textwidth]{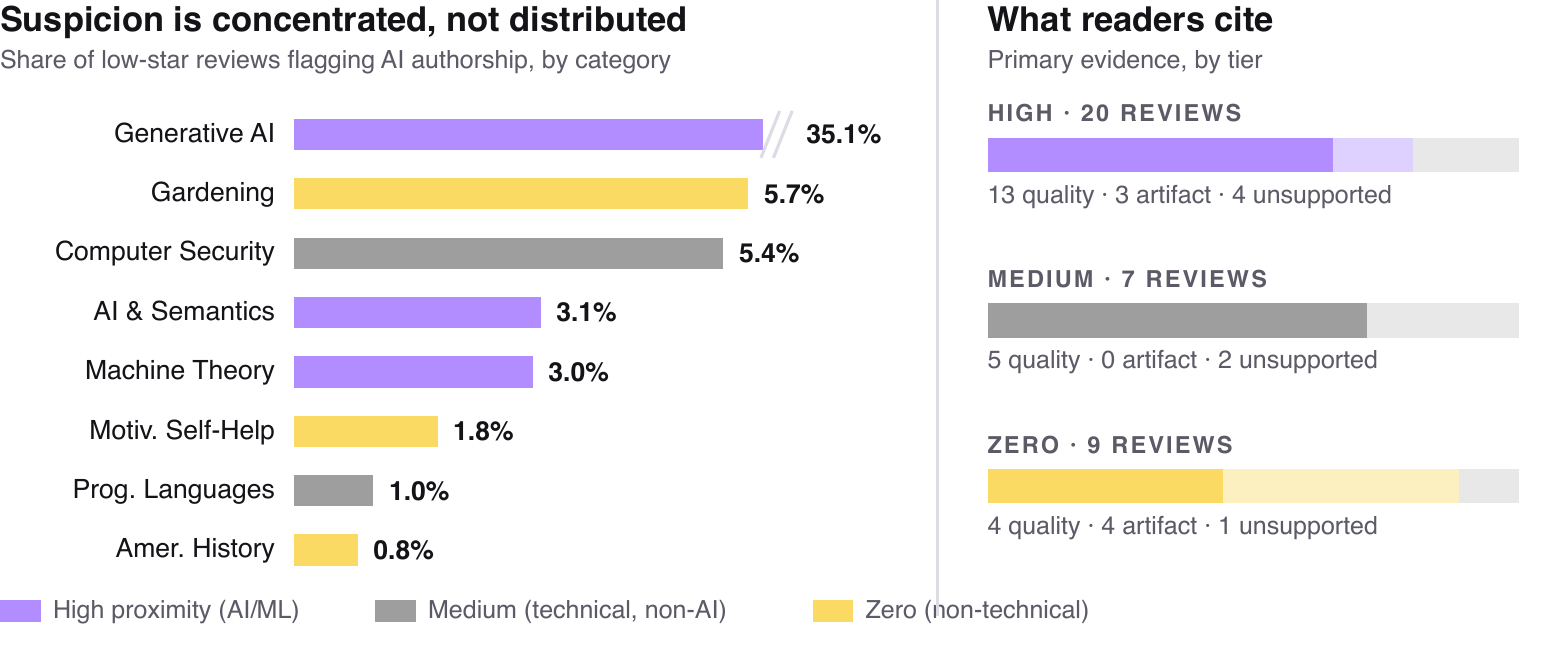}
\caption{AI authorship suspicion is concentrated rather than distributed. Left: the share of low-star reviews flagging AI authorship in each of the \nCategories{} categories, colored by topical proximity to AI. Generative AI (\genAiFlagRate{}, 13 of 37 reviews) is \genAiToNextRatio{} higher than any other category, and the axis is broken above 6\% so the remaining categories stay legible. Gardening, with zero topical proximity, shows the second-highest rate (\flagRateGardening{}). Right: the primary evidence readers cite, by proximity tier. Production-evidence heuristics (D6, D3) make up 4 of 9 coded reviews in zero-proximity categories against 3 of 20 in high-proximity ones.}
\label{fig:ai-perception}
\Description{Two panels. The left panel is a horizontal bar chart of eight categories ordered by flag rate, each bar shaded by proximity tier: Generative AI 35.1 percent, Gardening 5.7, Computer Security 5.4, AI and Semantics 3.1, Machine Theory 3.0, Motivational Self-Help 1.8, Programming Languages 1.0, American History 0.8. The Generative AI bar crosses an axis break above 6 percent, so it is drawn shorter than its value would otherwise place it. A legend maps three shades to high proximity (AI/ML), medium (technical, non-AI) and zero (non-technical). The right panel lists the primary evidence coded in each tier: high proximity, 20 reviews, 13 quality, 3 artifact, 4 unsupported; medium, 7 reviews, 5 quality, 0 artifact, 2 unsupported; zero, 9 reviews, 4 quality, 4 artifact, 1 unsupported.}
\end{figure*}

Book marketplaces rely on quality signals (star ratings, review counts, bestseller rank) built for an era when content was \textit{human-authored}. As AI becomes part of how text content is authored (supply), new questions arise on its impact on readers (demand).
% Supply documented; platform response inadequate; reviews as the window
Recent research has documented trends in the supply side of this shift. For example, monthly book releases tripled between 2022 and late 2025, and AI-containing titles now top half of new releases~\cite{reimers2026ai}. Platforms have responded with measures such as voluntary AI use disclosure~\cite{amazon2023kdp}. 
% But disclosure is self-reported, unverified, and not surfaced to consumers at purchase: readers are left to judge for themselves. 
However, how readers respond, the demand side, is far less understood. Reviews offer a unique window into this phenomenon: they are among the few places where readers articulate, unprompted and at length, their experience with content they suspect is AI-generated. We examine this in reviews of bestselling books across categories varying in topical proximity to AI.

% Research questions
Interpreting AI suspicion requires first situating it within readers' broader complaints. We address three research questions:
\begin{description}
    \item[RQ1] What are the dominant dimensions of reader complaints in bestselling books, and how do they vary across categories?
    \item[RQ2] To what extent do readers perceive AI-generated authorship in bestselling books, how does this perception manifest in their language, and what book-level signals (publisher type, verified purchase rates, review verbosity) predict it?
    \item[RQ3] Does AI authorship suspicion generalize beyond AI/ML books, or is it specific to categories where AI is the subject matter?
\end{description}

% What we did
To address these questions, we conducted an observational study of reader reviews on Amazon bestseller lists. We collected \nReviews{} reviews across \nBooks{} books in \nCategories{} categories selected to vary on two properties of the books themselves: whether the subject matter concerns AI, and whether it is technical. Three AI/ML categories (Generative AI, AI \& Semantics, Machine Theory) represent {\textcolor{tierHigh}{\rule[-1pt]{3.5pt}{8pt}}}\,high proximity. Two technical non-AI categories (Computer Security, Programming Languages) represent {\textcolor{tierMed}{\rule[-1pt]{3.5pt}{8pt}}}\,medium proximity. Three non-technical categories (Motivational Self-Help, American History, Gardening) represent {\textcolor{tierZero}{\rule[-1pt]{3.5pt}{8pt}}}\,zero proximity. This three-tier design separates the AI-topic factor from the technical-subject-matter factor. Each review was scored on continuous dimensions using an LLM-based scoring pipeline validated via inter-model agreement (ICC = \iccAI{} on the AI authorship dimension), measuring the prominence of AI authorship suspicion alongside content quality, accuracy, presentation, and meta issues. AI suspicion analyses focus on the \nLowStar{} low-star (1--3 star) reviews where the construct is articulated; high-star reviews are retained as a contrast set for book-level signals like verified-purchase rates and review verbosity.

% Key findings
Our analysis yields three main findings:

\begin{enumerate}
    \item \textbf{AI authorship suspicion follows a bimodal pattern.} Overall, \overallFlagRate{} of critical reviews flag AI authorship (score $\geq 5$ on a 1--10 scale). This suspicion is concentrated in Generative AI books (\genAiFlagRate{}) but does not exceed 6\% in any other category (Figure~\ref{fig:ai-perception}). Yet AI suspicion appears in every category we examined, including non-technical ones: the Gardening category, with zero topical proximity to AI, shows the second-highest flag rate (\flagRateGardening{}), driven by concrete artifacts (phantom authors, AI-generated imagery, suspected ghostwriting). Across all books, technical non-AI categories sit within the same range as non-technical ones: Computer Security (\flagRateComputerSecurity{}) and Programming Languages (\flagRateProgLanguages{}). This contrast is sensitive to publication recency, and we report a recency-matched comparison alongside it (Section~\ref{sec:proximity}).

    \item \textbf{AI suspicion co-occurs with content and presentation failures, not meta issues.} Across all categories, content shallowness and meta issues are the primary reader complaints. When readers do suspect AI authorship, their reviews complain more about shallow content ($d = \crosstabCQd{}$) and poor presentation ($d = \crosstabPRESd{}$), and less about misleading titles or formatting ($d = \crosstabMETAd{}$), than other critical reviews do.

    \item \textbf{AI suspicion manifests in two distinct patterns.} (i) reviewers use ``AI-generated'' as shorthand for generic or shallow writing -- a quality complaint rather than a literal authorship claim. (ii) multiple independent reviewers flag the same book with specific evidence (``filled with GPT responses,'' ``author doesn't exist''). This second pattern is concentrated in a small number of self-published Generative AI titles and represents a stronger signal of likely AI involvement.
\end{enumerate}

% Contributions
This paper makes two contributions:

\begin{enumerate}
    \item To our knowledge, the first empirical study of how organic marketplace reviews express AI authorship suspicion. We document a bimodal pattern (a Generative AI hotspot alongside a low but consistent presence in every other category), the complaint profile that accompanies suspicion, and the heuristics readers cite when flagging it. These measurements date the pattern, so a later replication can test whether the hotspot spreads and the baseline rises. We propose two mechanisms operating in parallel: topical concentration and direct artifact detection.

    \item Release of the corpus, scoring rubric, and analysis code under open licenses. The dimensionally scored reviews provide labeled training data for the AI authorship suspicion construct, supporting small deployable classifiers where frontier-LLM scoring is cost-prohibitive at platform scale; held-out splits, baselines, and a distilled scorer will be released with the corpus.
\end{enumerate}

\section{Related Work}
\label{sec:related-work}

We situate this work in research on AI text detection, the effects of explicit AI disclosure, and the growth of AI-generated content in book marketplaces, then identify the mechanisms that could plausibly produce reader suspicion where authorship is unverifiable.

\subsection{Human Detection of AI-Generated Text}
\label{sec:rw-detection}

People are poor at telling AI-written text from human-written text. Detection accuracy sits at 50--52\% for AI-generated self-presentations across six experiments with 4,600 participants~\cite{jakesch2023heuristics}, essentially coin-flip performance. Training does not close the gap: untrained evaluators cannot distinguish GPT-3 output from human writing, and training interventions lift accuracy only to about 55\%~\cite{clark2021human}. Nor is the failure confined to prose. In poetry, readers not only misidentify AI-generated work (46.6\% accuracy) but rate it \textit{higher} than human-written poetry on rhythm and beauty~\cite{porter2024poetry}. Detection ability is not uniform across readers, however. Annotators who use LLMs daily for writing identify AI-generated non-fiction with high accuracy and justify their judgments with specific textual cues~\cite{russell2025detectors}, a population that overlaps with the readership of the AI/ML books we sample.

These findings frame our study. Detection ability appears to depend on the reader, so the flags we observe in Amazon reviews could reflect genuine recognition by AI-familiar readers, misattribution by readers who cannot in fact detect, or both. Which is operating, and where, is an open question.

\subsection{AI Disclosure and Trust}
\label{sec:rw-disclosure}

\textit{Knowing} content is AI-generated consistently reduces how people evaluate it, even when the content is objectively equivalent. Across 13 preregistered experiments with more than 3,000 participants, disclosure reduced trust through reduced perceived legitimacy, regardless of whether participants already suspected AI use~\cite{schilke2025transparency}. The penalty extends beyond text: across 565 participants, identical paintings presented as ``AI-generated'' rather than ``human-created'' were rated lower on liking, beauty, novelty, and meaning~\cite{ragot2020ai}. It also predates LLMs. News articles declared human-written were rated more favorably than identical computer-written versions, even when the latter scored higher on credibility in blind evaluation~\cite{graefe2018perception}.

The penalty is also specific to settings where provenance is uncertain. Trust penalties for AI-written content emerge in mixed environments where some content is believed to be AI-generated and some is not, an effect termed the ``Replicant Effect''~\cite{jakesch2019replicant}. Amazon is exactly such an environment, where suspicion can attach to any individual title without verifiable provenance.

These studies examine \textit{explicit} disclosure, where participants are told content is AI-generated. Our setting has no such label: readers form suspicion through the pathways we consider in Section~\ref{sec:rw-mechanisms}.

\subsection{AI-Generated Content in Marketplaces}
\label{sec:rw-marketplace}

Books are an information good where pre-purchase quality assessment is difficult~\cite{akerlof1970lemons, waldfogel2018selfpub}; AI-generated content is a further challenge to this assessment. Evidence of AI-generated books on Amazon is growing. Monthly releases tripled between 2022 and late 2025, with AI-containing titles topping half of 2025 releases~\cite{reimers2026ai}; industry groups have raised concerns about the scale of AI-authored self-published titles~\cite{nyt2024aibooks}. Prevalence is hard to establish precisely, since estimates depend on detectors whose accuracy varies sharply with how the text was produced (Section~\ref{sec:limitations}). Concurrent work measures the same market from the supply side: Chakrabarty et al.~\cite{chakrabarty2026flood} apply full-text AI detection to 14,419 self-published genre-fiction titles matched to Amazon sales records and find that books with substantial detected AI text reach commercial scale and take top-rank positions from books with no detected AI text, none of them disclosing AI use. That work measures what is produced and sold; we measure what readers say, in a disjoint corpus of technical and trade non-fiction.

\subsection{Mechanisms of AI Authorship Attribution}
\label{sec:rw-mechanisms}

Several accounts could plausibly explain reader suspicion of AI authorship in the absence of explicit disclosure. \textit{Cognitive priming} offers one account. The construct accessibility model~\cite{higgins1977accessibility, higgins1996knowledge} holds that recently activated concepts influence subsequent judgments when the concept is applicable to the stimulus. Reading a book about AI activates ``AI-generated content'' as an applicable category, which the availability heuristic~\cite{tversky1973availability} can render more accessible when readers encounter quality concerns; confirmation bias~\cite{nickerson1998confirmation} may then sustain the interpretation. Prior work tested whether a text's topic changes the penalty for labeled AI authorship and mostly found it does not: topic shifted judgments of author competence and left content quality and sharing intention unchanged~\cite{proksch2024topic}. We ask whether topic alone, with no label at all, is enough to draw suspicion.

\textit{Direct artifact detection} offers a different account. The detection literature in Section~\ref{sec:rw-detection} concerns prose-style assessment, where humans perform near chance. But some AI-generated content carries visible production signatures that do not require stylistic discrimination: phantom or inconsistent author identities, AI-generated cover or interior imagery, and verbatim prompt-and-response transcripts. This pathway operates regardless of subject matter.

\textit{Base-rate effects} offer a competing account of the same pattern. If some categories genuinely contain more AI-generated content, suspicion there is warranted rather than amplified, and would reach readers through the artifact detection above. We cannot establish category-level prevalence for our corpus, since detector sensitivity varies with how the text was produced (Section~\ref{sec:limitations}), but book-level production signals offer a partial proxy (Section~\ref{sec:rq3}).

These mechanisms are not mutually exclusive, and distinguishing them requires observing suspicion where it arises naturally rather than where it is induced. Marketplace reviews provide that setting: readers articulate suspicion unprompted, across categories that vary in topical proximity to AI, letting us test whether suspicion tracks subject matter or production artifacts.

\section{Methodology}
\label{sec:methodology}

%% Declared at the head of the section so it floats to the top of the page the
%% Methodology opens on. figure* reaches page tops only; verify the landing page
%% in the built PDF, never the declaration.
\begin{figure*}[t]
\centering
\includegraphics[width=\textwidth]{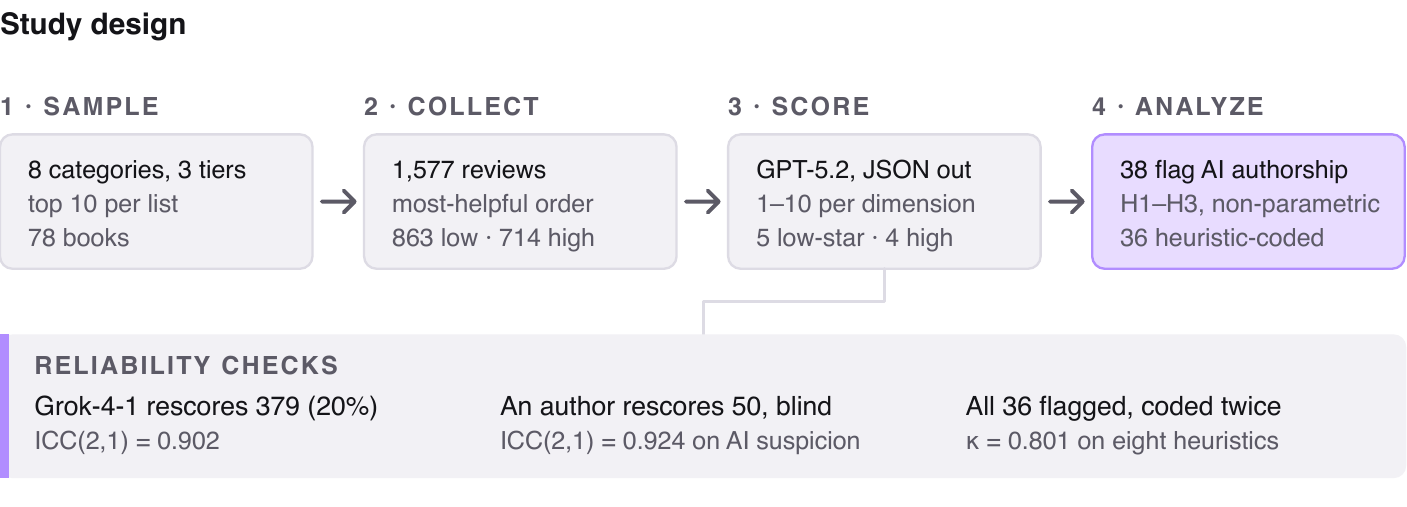}
\caption{Study design. The eight subcategories are listed individually in Table~\ref{tab:dataset} and the three proximity tiers they span are described in Section~\ref{sec:categories}. The reliability checks cover two distinct annotation tasks: the dimension scores, rescored by a second model and by an author, and the qualitative heuristic coding, performed twice.}
\label{fig:method}
\Description{A four-stage pipeline running left to right in boxes, with a reliability panel beneath it. Stage 1, sample: 8 categories across 3 tiers, top 10 per bestseller list, 78 books. Stage 2, collect: 1,577 reviews taken in most-helpful order, 863 low-star and 714 high-star. Stage 3, score: GPT-5.2 with JSON output, 1 to 10 per dimension, five dimensions for low-star reviews and four for high-star. Stage 4, analyze: 38 reviews flag AI authorship, hypotheses H1 to H3 tested with non-parametric methods, 36 reviews heuristic-coded. The panel below lists three reliability checks: Grok-4-1 rescores 379 reviews, 20 percent, giving ICC(2,1) of 0.902; an author rescores 50 reviews blind, giving ICC(2,1) of 0.924 on AI suspicion; and all 36 flagged reviews are coded twice, giving a kappa of 0.801 across the eight heuristics.}
\end{figure*}

This section describes our study design for addressing RQ1--RQ3 (Section~\ref{sec:introduction}). We collected reviews from Amazon bestseller pages across eight subcategories spanning three proximity tiers, scored each review on five dimensions using a frontier LLM with structured output, and validated scoring through inter-model agreement and a stratified human subsample. Three hypotheses were tested using non-parametric methods on the full dataset.

\subsection{Category Selection}
\label{sec:categories}

We sampled books from eight Amazon bestseller subcategories, selected to vary on two properties of the books themselves: whether the subject matter concerns AI, and whether it is technical. The decomposition supports two comparisons: high- vs.\ medium-proximity isolates the AI-topic factor while holding technical subject matter constant; medium- vs.\ zero-proximity isolates the technical factor with AI topic absent from both.

The three high-proximity categories sit under Computers \& Technology $\rightarrow$ AI \& Machine Learning: Artificial Intelligence \& Semantics mixes popular and technical titles; Machine Theory is the most academic of the three and is dominated by established publishers; Generative AI is the newest subcategory and carries the highest proportion of self-published titles. The two medium-proximity categories are technical but not about AI: Computer Security consists largely of certification-preparation titles, and Programming Languages is an established textbook market with substantial independent participation. The three zero-proximity categories have neither AI nor technical subject matter, and were chosen to vary in self-publishing penetration: American History is dominated by Big Five publishers, Motivational Self-Help carries more independent participation, and Gardening, with substantial self-publishing, serves as a second non-technical comparison.

We enumerated candidate subcategories, counted independently published against established-publisher titles on each bestseller page, and selected subcategories to maximize variation across the proximity and publishing dimensions. Table~\ref{tab:dataset} lists each subcategory with its Amazon browse-node identifier, and Appendix~\ref{app:category-selection} reports the full category discovery results.

\subsection{Sampling Protocol}
\label{sec:sampling}

\subsubsection{Book Selection.}
We took the top 10 books from each bestseller page at the time of collection. For the three zero-proximity categories we over-collected the top 30 and applied a recency filter (publication date $\geq$ January 2023) before taking the top 10 by rank, because non-technical bestseller lists are dominated by evergreen titles that predate the LLM era. The AI/ML categories did not need this filter: the Generative AI subcategory was created after ChatGPT's launch, and all three are dominated by recent publications. The medium-proximity categories were taken by rank alone, and are the least recent of the three tiers as a result (\recHighBookPct{}, \recMedBookPct{} and \recZeroBookPct{} of high-, medium- and zero-proximity books were published in 2023 or later). Because a book published before ChatGPT cannot contain LLM-generated text, part of the medium tier can draw only ambient suspicion, which lowers its flag rate relative to the other two tiers and understates its distance from the non-technical baseline. Section~\ref{sec:proximity} reports both the full-sample and the recency-matched comparison.

\subsubsection{Publisher Classification.}
Each of the \nPublishers{} distinct publishers in the corpus was classified as self-published or established using the two-model protocol applied elsewhere in this study (GPT-5.2 primary, Grok-4-1 validation; \publisherAgreement{} raw agreement). The \nPublisherAdjudicated{} cases the protocol flagged for review, where the models disagreed or either returned an uncertain or low-confidence label, were adjudicated by hand against web evidence. The construct is editorial gatekeeping rather than legal form, so author-owned imprints count as self-published. \nPublisherUnresolved{} books whose publisher could not be established with confidence are excluded from publisher analyses. The full classification, including per-publisher evidence, will be released with the dataset.

\subsubsection{Review Collection.}
For each book we collected reviews via browser automation (Playwright-based accessibility tree snapshots): the first two pages of critical reviews (Amazon's 1--3 star filter), the first page of 1-star reviews, and the first page of 5-star reviews. The separate 1-star draw ensures the harshest reviews are well represented, which a helpfulness-sorted critical page can under-surface. The two low-star draws overlap and are deduplicated when snapshots are parsed, and the resulting analysis set is \oneStarShare{} 1-star by design. Low-star reviews carry the complaint signals we study, so we collect them at greater depth; high-star reviews serve as a contrast set for book-level measures such as verified-purchase rate and review verbosity, for which one page suffices.

We used Amazon's default sort (most helpful), so our sample captures the reviews other readers have voted most useful, which are also the reviews most likely to influence prospective buyers. Combined with the oversampling of critical reviews, this means our data characterizes the \textit{most visible criticism} rather than the overall reader experience.

\subsubsection{Dataset Summary.}
Table~\ref{tab:dataset} summarizes the collected data. Collection spanned February 28 to March 7, 2026.

\begin{table}[h]
\caption{Dataset summary by category and proximity tier. Node is the Amazon browse-node identifier, which is stable when a subcategory is renamed. Low-star = 1--3 star reviews; high-star = 5-star reviews.}
\label{tab:dataset}
\resizebox{\narrowwidth}{!}{%
\begin{tabular}{lllrrr}
\toprule
Category & Proximity & Node & Books & Low-star & High-star \\
\midrule
AI \& Semantics & High & 280291 & 10 & 131 & 93 \\
Machine Theory & High & 280292 & 9 & 99 & 84 \\
Generative AI & High & 211759007011 & 9 & 37 & 81 \\
\midrule
Amer. History & Zero & 4808 & 10 & 130 & 95 \\
Motiv. Self-Help & Zero & 4744 & 10 & 110 & 97 \\
\midrule
Computer Security & Medium & 377866011 & 10 & 111 & 87 \\
Prog. Languages & Medium & 3839 & 10 & 104 & 92 \\
Gardening & Zero & 48 & 10 & 141 & 85 \\
\midrule
\textbf{Total} & & & \textbf{78} & \textbf{863} & \textbf{714} \\
\bottomrule
\end{tabular}}
\end{table}

Reviews were deduplicated by matching the first 100 characters of review body within each book (the \texttt{body\_prefix} key in the released dataset); 38 duplicate reviews and 2 duplicate book entries were removed. The 2 duplicate books are titles that chart on two of our sampled bestseller lists at once; each is retained once and assigned to a single category, so per-category counts are 9 rather than 10 for Machine Theory and Generative AI. All hypothesis tests use the full \nCategories{}-category dataset (\nBooks{} books, \nLowStar{} low-star reviews) unless otherwise noted. Book-level publisher comparisons use $N = \hFourIndieN{}$ self-published and $N = \hFourEstabN{}$ established-publisher books; \nPublisherExcluded{} books are excluded, \nPublisherUnresolved{} for an unresolved publisher and \nNoLowStarClassified{} for returning no low-star reviews and therefore no AI perception score. \nNoLowStarBooks{} books in total returned no low-star reviews and are excluded from book-level flag-rate comparisons rather than counted as unflagged.

\subsection{Dimensional Review Scoring}
\label{sec:scoring}

\subsubsection{Scoring Approach.}
Each review was scored on continuous 1--10 dimensions using a frontier LLM (OpenAI GPT-5.2) with structured JSON output. Scores indicate how prominently each dimension features in the review (1 = absent, 10 = dominant theme), with a one-sentence rationale per score. This approach follows established practice for using LLMs as annotation tools for subjective constructs~\cite{gilardi2023chatgpt, zheng2023judging, rathje2024gpt, tornberg2024bestpractices}.

\textbf{Low-star review dimensions} (5): content quality, accuracy, AI generation perception, presentation, meta issues.

\textbf{High-star review dimensions} (4): content quality, practical value, presentation, logistical only. See Appendix~\ref{app:scoring-prompt} for the full scoring prompt.

\subsubsection{AI Generation Perception: Prompt Design.}
The AI generation perception dimension required careful prompt engineering to distinguish AI-as-subject from AI-as-authorship. The scoring prompt includes the instruction: \textit{``Only score high if the reviewer suspects the BOOK CONTENT was AI-generated. Do NOT score high just because the book's SUBJECT MATTER is about AI/ChatGPT/LLMs.''} We validated this distinction by confirming that books about AI (e.g., \textit{The Coming Wave}) score low on this dimension when reviewers do not suspect AI authorship. The dimension measures what a reviewer asserts, not whether the book was in fact AI-generated. Suspicion is consequential on its own terms: it moves purchases and attaches to named authors regardless of its accuracy, and it is what a prospective buyer encounters on the page.

\subsubsection{Inter-Model Validation.}
\label{sec:validation}
To assess the reliability of LLM-based scoring, a stratified \imSamplePct{} sample (\imNSampled{} reviews) was independently scored by a second frontier model (xAI Grok-4-1), given identical prompts and scoring schemas with no access to the first model's outputs. Across all dimensions ($N = \imNPairs{}$ score pairs) the models reach an intraclass correlation of ICC(2,1) = \iccOverall{}, weighted $\kappa$ = \imKappaOverall{}, Pearson $r$ = \imPearsonOverall{}, and within-1 agreement of \imWithinOne{}. Agreement is highest on AI generation perception (ICC(2,1) = \iccAI{}, weighted $\kappa$ = \imKappaAI{}, exact agreement = \imExactAI{}) and lowest on logistical-only complaints (\imIccMin{}, a low-variance dimension); the lowest of the five low-star dimensions is \imLowStarIccMin{}. Appendix~\ref{app:validation} reports full agreement tables.

\subsubsection{Human Validation.}
\label{sec:human-validation}
To complement the inter-model validation, a stratified subsample of \humanValN{} reviews (\humanValFlaggedN{} from the AI-flagged stratum and \humanValNonFlaggedN{} randomly sampled from the non-flagged strata) was independently scored by an author, blind to the LLM scores, dimension rationales, and the review's stratum or category assignment.

Following Koo \& Li~\cite{koo2016icc}, we interpret ICC values based on the 95\% confidence interval rather than the point estimate alone (cutoffs: $<.50$ poor, .50--.75 moderate, .75--.90 good, $>.90$ excellent). For the primary AI authorship suspicion construct, human-LLM agreement was good-to-excellent: ICC(2,1) = \humanIccAI{}, 95\% CI [\humanIccAIciLo{}, \humanIccAIciHi{}]. For the binary AI flag classification (score $\geq 5$), Cohen's $\kappa$ = \humanAiKappa{} with \humanAiAgreement{} raw agreement (\humanAiTP{} true positives, \humanAiTN{} true negatives, \humanAiFP{} false positive, \humanAiFN{} false negatives). Content quality showed moderate-to-good agreement (ICC = \humanIccCQ{}, 95\% CI [\humanIccCQciLo{}, \humanIccCQciHi{}]). Supporting dimensions showed wider CIs spanning poor-to-good interpretations (accuracy ICC = \humanIccAcc{} [\humanIccAccciLo{}, \humanIccAccciHi{}], presentation ICC = \humanIccPres{} [\humanIccPresciLo{}, \humanIccPresciHi{}], meta issues ICC = \humanIccMeta{} [\humanIccMetaciLo{}, \humanIccMetaciHi{}]), reflecting both construct boundary fuzziness on these dimensions and the modest sample size; we discuss this further in the limitations.

\subsection{Statistical Methods}
\label{sec:stats}

All tests are non-parametric because dimension scores are ordinal (1--10) and fail Shapiro-Wilk normality tests across all groups. We use the Kruskal-Wallis H-test with Bonferroni-corrected Dunn's post-hoc tests for multi-group comparisons, Mann-Whitney U for two-group comparisons, Spearman rank correlation for associations, and the chi-square test of independence for flag-rate comparisons. Book-level analyses use per-book averages to avoid pseudoreplication from treating reviews within the same book as independent observations.

Effect sizes are reported alongside all $p$-values: epsilon-squared for Kruskal-Wallis, Cohen's $d$ (computed on raw scores) for Mann-Whitney comparisons as a widely understood benchmark, rank-biserial correlation $r_{rb}$ as the appropriate ordinal effect size for Mann-Whitney tests, and Cram\'er's $V$ for chi-square. 95\% confidence intervals are computed via bootstrap (10,000 resamples, seed = 42). Significance threshold $\alpha = 0.05$. All analysis code and random seeds are included in the reproducibility package; the full benchmark structure (held-out splits, official tasks, baseline scorer) will be released with the corpus.

\subsection{Hypotheses}
\label{sec:hypotheses}

Our analyses test three hypotheses derived from the research questions:

\begin{description}
    \item[H1 (Concentration)] AI authorship suspicion is higher in AI/ML categories than in non-technical categories, and is concentrated in the Generative AI subcategory rather than distributed across AI/ML books generally. We test this with both continuous scores (Kruskal-Wallis with Dunn's post-hoc) and binary flag rates (chi-square), and include a planned contrast comparing AI/ML categories against non-AI categories (Mann-Whitney U, one-sided).

    \item[H2 (Co-occurrence)] Reviews expressing AI authorship suspicion show elevated content quality and presentation complaints, and reduced meta/logistical complaints, relative to non-suspicious low-star reviews. We test this by comparing dimension profiles of AI-flagged versus non-flagged reviews (Mann-Whitney U per dimension).

    \item[H3 (Metadata)] Book-level metadata signals -- publisher type (independent vs.\ established), 5-star verified purchase rate, and 5-star review verbosity -- predict AI authorship suspicion. These signals test whether suspicion tracks book-level production characteristics in addition to category-level effects (Mann-Whitney U, two-sided, for publisher type; Spearman correlation for the continuous signals). Publisher type is assigned from the publisher classification described in Section~\ref{sec:sampling}; books whose publisher could not be classified are excluded from this test.
\end{description}

\subsection{Qualitative Heuristic Coding}
\label{sec:heuristic-coding}

To characterize the evidence readers cite when suspecting AI authorship, an author developed a codebook through iterative open coding. An initial pass over the flagged reviews identified recurring justification patterns, which were consolidated into eight detection heuristics (D0--D7):

\begin{description}
    \item[D0] Unsupported assertion -- reviewer states ``this is AI'' without citing evidence.
    \item[D1] Prose style recognition -- reviewer identifies stylistic markers associated with LLM output (e.g., em dashes, repetitive structure).
    \item[D2] Shallow/generic content -- reviewer claims the content lacks depth or originality.
    \item[D3] Phantom/unverifiable author -- reviewer questions whether the author exists.
    \item[D4] Poor writing quality -- reviewer cites grammar, coherence, or editing failures.
    \item[D5] Factual errors -- reviewer identifies specific incorrect claims.
    \item[D6] AI output visible in product -- reviewer identifies concrete AI artifacts (e.g., AI-generated images, ChatGPT-style formatting).
    \item[D7] Style deviation from known author -- reviewer notes a departure from the author's established style.
\end{description}

Each AI-flagged review was assigned a single primary heuristic code based on the full review text (title plus body), not just the portion referencing AI, together with any secondary heuristics the review also invoked. Co-occurrence counts reported in the results use primary and secondary codes together. \nHeuristicExcluded{} of the \nFlagged{} flagged reviews were set aside during coding, both coders having found no AI-authorship claim in them, leaving \nHeuristicCoded{} coded reviews. Coding was performed by GPT-5.2 and independently repeated by Grok-4-1, each given the complete review and codebook with no access to the other's output; inter-model agreement was $\kappa = \heuristicKappa{}$, with exact agreement on \heuristicExactN{} of \nHeuristicCoded{} reviews (\heuristicExactPct{}). See Appendix~\ref{app:qualcoding} for full agreement details.

\section{Results}
\label{sec:results}

RQ1 (complaint landscape) is addressed descriptively. RQ2 (AI suspicion extent, manifestation, and metadata predictors) maps to three hypotheses: H1 (concentration), H2 (co-occurrence), and H3 (metadata). RQ3 (generalization beyond AI/ML) is addressed as a follow-up proximity gradient analysis.

\subsection{RQ1: Complaint Landscape}
\label{sec:rq1}

Across \nLowStar{} low-star reviews in \nCategories{} categories, the dominant complaint dimensions by mean score are: meta issues (\dimMeanMetaIssues{}/10), content quality (\dimMeanContentQuality{}/10), presentation (\dimMeanPresentation{}/10), accuracy (\dimMeanAccuracy{}/10), and AI generation perception (\dimMeanAiPerception{}/10). Content shallowness and meta issues (misleading titles, wrong audience, formatting problems) are the primary reader complaints across all categories. AI authorship suspicion is the least prominent dimension.

Content quality scores differ significantly across categories ($H(\hThreeContentDf{}) = \hThreeContentH{}$, $p < .001$, $\varepsilon^2 = \hThreeContentEps{}$, \hThreeContentEffect{} effect). Eight pairwise comparisons survive Bonferroni correction: AI/ML categories (AI \& Semantics, Generative AI) and two non-technical categories (Motivational Self-Help, American History) receive higher content quality complaints than technical non-AI categories (Computer Security, Programming Languages) and Gardening. The pattern does not align cleanly with proximity tier: Gardening, despite being non-technical, sits with the technical categories at the low end, possibly reflecting that gardening books rely more on visual and practical content than on extended prose. Content shallowness is a widespread concern, but its intensity varies by genre.

\subsection{RQ2: AI Authorship Suspicion}
\label{sec:rq2}

\subsubsection{H1 -- Concentration.}

AI generation perception scores differ significantly across categories ($H(\hOneDf{}) = \hOneH{}$, $p < .001$, $\varepsilon^2 = \hOneEpsSq{}$, \hOneEffect{} effect). Post-hoc Dunn's tests with Bonferroni correction show that Generative AI ($M = \genAiMean{}$) differs significantly from all other categories ($p < .001$ for each); no other pairwise comparisons reach significance. AI suspicion is elevated in one subcategory, not distributed across AI/ML categories generally.

We define a review as ``AI-flagged'' when its AI generation perception score $\geq 5$ (moderate signal or above). Flag rates differ significantly across categories (Table~\ref{tab:flagrates}; $\chi^2(\hEightDf{}) = \hEightChi{}$, $p < .001$, Cram\'er's $V = \hEightV{}$, \hEightEffect{} effect):

\begin{table}[h]
\caption{Reviews flagging AI authorship, by category. $N$ is the number of low-star (1--3 star) reviews in the category; a review is flagged when its AI generation perception score is 5 or higher. Categories are ordered by flag rate.}
\label{tab:flagrates}
\resizebox{\narrowwidth}{!}{%
\begin{tabular}{lrrrr}
\toprule
Category & $N$ & Flagged & Not flagged & Flag rate \\
\midrule
Generative AI & 37 & 13 & 24 & 35.1\% \\
Gardening & 141 & 8 & 133 & 5.7\% \\
Computer Security & 111 & 6 & 105 & 5.4\% \\
AI \& Semantics & 131 & 4 & 127 & 3.1\% \\
Machine Theory & 99 & 3 & 96 & 3.0\% \\
Motiv. Self-Help & 110 & 2 & 108 & 1.8\% \\
Prog. Languages & 104 & 1 & 103 & 1.0\% \\
Amer. History & 130 & 1 & 129 & 0.8\% \\
\bottomrule
\end{tabular}}
\end{table}

Overall, \overallFlagRate{} of critical reviews (\nFlagged{}/\nLowStar{}) flag AI authorship. The Generative AI rate is \genAiToNextRatio{} higher than any other category (\nextHighestCat{} at \nextHighestRate{}). The planned contrast shows that AI/ML categories ($M = \hTwoAiM{}$, $N = \hTwoAiN{}$) have significantly higher AI perception than non-AI categories ($M = \hTwoNonAiM{}$, $N = \hTwoNonAiN{}$; $U = \hTwoU{}$, $p = \hTwoP{}$, $d = \hTwoD{}$, small effect).

The flag rates reveal a bimodal pattern: a dramatic concentration in Generative AI (\genAiFlagRate{}) alongside a low but consistent presence in every other category we examined (all under 6\%). Gardening (\flagRateGardening{}) and Computer Security (\flagRateComputerSecurity{}) show the highest non-Generative-AI rates despite their distance from the AI subject matter. We return to this cross-category presence in Section~\ref{sec:proximity}.

\subsubsection{Book-Level Concentration.}
Within Generative AI, AI suspicion is driven by specific titles: \textit{The ChatGPT Millionaire} (\mostFlaggedFlagPct{} flag rate, \mostFlaggedFlagged{}/\mostFlaggedLowStar{} reviews) and \textit{Build Passive Income with AI} (\nextFlaggedPct{} flag rate, \nextFlaggedFlagged{}/\nextFlaggedLowStar{} reviews). Both are self-published. Three titles account for all 13 flagged reviews in the category.

\subsubsection{Two Suspicion Patterns.}
We observe two distinct forms of AI authorship suspicion:

\textbf{Pattern A -- Ambient suspicion.} In AI \& Semantics and Machine Theory, isolated reviewers use ``AI-generated'' as criticism of writing they find generic or repetitive. These are single-reviewer flags without corroboration, suggesting ``AI-generated'' is entering the critical vocabulary as shorthand for formulaic writing. \preChatGptBookFlags{} flagged reviews make this explicit by targeting books that predate ChatGPT: a 2018 title reviewed in 2024 (``most of this book is either AI generated'') and a 2019 title reviewed in 2025 (``read like copied from ChatGPT''). Neither book can contain LLM-generated text, so these accusations reinterpret perceived writing quality rather than detect provenance.

\textbf{Pattern B -- Corroborated suspicion.} A small number of individual books receive AI generation accusations from multiple independent reviewers. Two or more flags fall on \nTwoFlagBooks{} books. Of these, \nMultiFlagBooks{} draw three or more: two self-published Generative AI titles and a CompTIA certification guide from an established technical publisher. Convergence alone does not establish corroboration, since the reviewers of a book may converge on an unsupported assertion. Of those \nTwoFlagBooks{}, \nProdMultiBooks{} pair convergence with production evidence, including two Gardening titles whose reviewers each cite a phantom author or AI-generated images.

The distinction matters for platform design: ambient suspicion carries high false-positive risk, while corroborated suspicion aggregates independent reader judgments into a stronger signal. This pattern distinction is orthogonal to the bimodal mechanism observed in flag rates (Section~\ref{sec:proximity}): ambient and corroborated suspicion can each arise from either topical concentration in AI-related categories or from artifact detection in any category.

\subsubsection{What Evidence Do Readers Cite?}
\label{sec:qualitative}
When readers express suspicion of AI authorship, what are they actually pointing to? Content quality problems? Stylistic tells? Or do they simply assert it without evidence? Using the heuristic codebook described in Section~\ref{sec:heuristic-coding}, we coded \nHeuristicCoded{} of the \nFlagged{} AI-flagged reviews. The remaining \nHeuristicExcluded{}, both from Motivational Self-Help, express dissatisfaction with writing quality without referencing AI authorship at all, and are scorer false positives rather than codeable accusations (\nHeuristicExcluded{}/\nFlagged{}; see Section~\ref{sec:limitations}).

\begin{table}[h]
\caption{Primary heuristic assigned to each of the \nHeuristicCoded{} AI-flagged reviews that could be coded. Two of the \nFlagged{} flagged reviews make no AI authorship claim and are excluded (Section~\ref{sec:heuristic-coding}).}
\label{tab:heuristics}
\resizebox{\narrowwidth}{!}{%
\begin{tabular}{llrr}
\toprule
Code & Heuristic & Count & \% \\
\midrule
D1 & Prose style recognition & 8 & 22 \\
D0 & Unsupported assertion (no evidence) & 7 & 19 \\
D2 & Shallow/generic content & 7 & 19 \\
D4 & Poor writing quality & 4 & 11 \\
D6 & AI output visible in product & 4 & 11 \\
D3 & Phantom/unverifiable author & 3 & 8 \\
D5 & Factual errors & 2 & 6 \\
D7 & Style deviation from known author & 1 & 3 \\
\bottomrule
\end{tabular}}
\end{table}

\begin{table*}[t]
\caption{Representative review excerpts illustrating AI suspicion heuristics. Reviews are lightly edited for length; book titles abbreviated.}
\label{tab:examples}
\resizebox{\textwidth}{!}{%
\begin{tabular}{llp{0.6\textwidth}l}
\toprule
Heuristic & Category & Excerpt & Pattern \\
\midrule
D1: Prose style & AI \& Sem. & ``I recognize a lot of ChatGPT's signature grammar all throughout this book. The em dashes and three point sentences, combined with the book covering the same exact topic cyclically seems awfully suspicious.'' (\textit{The Coming Wave}) & Ambient \\
\addlinespace
D0: Unsupported assertion & Gardening & ``Feels like it was written by AI.'' (\textit{Martha Stewart's Gardening Handbook}) & Ambient \\
\addlinespace
D2: Shallow content & Gen. AI & ``Written for an audience of third graders\ldots\ The only chapter of substance\ldots\ There's not much real value in any of it\ldots\ Considering [the author] practically had ChatGPT write the book for him.'' (\textit{ChatGPT Millionaire}) & Corroborated \\
\addlinespace
D3: Phantom author & Gardening & ``This book was generated by artificial intelligence and the author doesn't exist. The information in the book is incorrect. There are photos of the wrong plants.'' (\textit{Herbal Synergies}) & Corroborated \\
\addlinespace
D6: AI artifacts & Gardening & ``I thought the little guy sitting on the tree stump was a carrot\ldots\ there are leaves growing out of its head\ldots\ small issues that weren't noticeable unless you were looking for them but would be very noticeable once you started coloring.'' (\textit{Tiny Gardens}) & Corroborated \\
\addlinespace
D7: Style deviation & Gardening & ``I've read many of her books and articles over the years and this doesn't read like her at all. In fact, it reads like a series of Google searches.'' (\textit{Martha Stewart's Gardening Handbook}) & Ambient \\
\bottomrule
\end{tabular}}
\end{table*}

Of the \nHeuristicCoded{} coded reviews, \dZeroCount{} are unsupported assertions (Table~\ref{tab:heuristics}): the reviewer states ``this is AI'' without citing evidence. These function as a label for dissatisfaction rather than a detection claim. Prose style recognition (\dOneCount{} reviews) and shallow content (\dTwoCount{}) are the most common evidence-based heuristics, and they co-occur in \dOneTwoCooccur{} of the \nHeuristicCoded{} coded reviews, as primary or secondary evidence. Reviewers who suspect AI authorship tend to point to a combination of generic prose and lack of substantive depth (Table~\ref{tab:examples} gives excerpts for each heuristic).

Heuristic usage differs by proximity tier, suggesting different pathways to suspicion. In high-proximity categories (AI/ML), readers cite shallow content (D2) and prose style (D1) -- quality-based heuristics that could reflect either genuine detection or topic-influenced interpretation. In zero-proximity categories (Gardening), readers cite visible AI artifacts (D6, e.g., AI-generated images in a coloring book) and phantom author concerns (D3, e.g., ``this author doesn't exist'') -- concrete production evidence that does not require subject-matter familiarity to notice. Quality-based heuristics in AI-topic categories and production-evidence heuristics elsewhere is a bimodal pattern consistent with two mechanisms operating in parallel.

\subsection{H2 -- The Complaint Profile of AI-Flagged Reviews}
\label{sec:rq2b}

\begin{figure}[t]
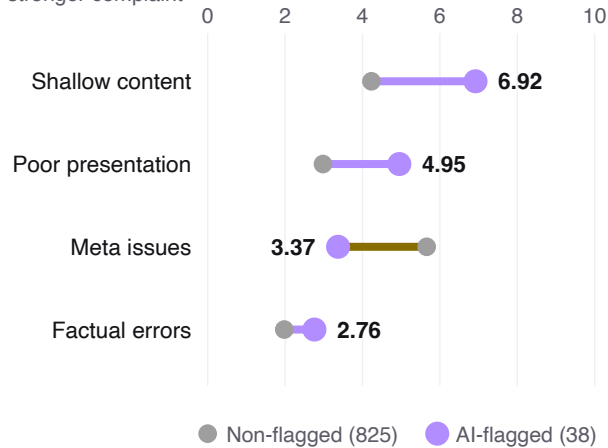

\centering
\profilefigure
\caption{Dimension means for AI-flagged ($n = \nFlagged{}$, score $\geq 5$) versus non-flagged ($n = \nNotFlagged{}$) low-star reviews. AI-flagged reviews score higher on shallow content, poor presentation and factual error complaints and \textit{lower} on meta issues, the one dimension that reverses. All comparisons use Mann-Whitney U, all $p < .005$.}
\label{fig:complaint-profile}
\Description{A connected dot plot with four rows, one per complaint dimension, on a scale from 0 to 10. Each row carries two markers joined by a line, the left marker for non-flagged reviews and the right for AI-flagged, except where the direction reverses. Shallow content runs from 4.23 non-flagged to 6.92 flagged; poor presentation from 2.98 to 4.95; factual errors from 1.98 to 2.76. Meta issues is the one row that runs the other way, from 5.66 non-flagged down to 3.37 flagged, so its flagged marker sits to the left. Only the flagged value is printed beside each row.}
\end{figure}

To understand what readers experience alongside AI suspicion, we compared the dimension profiles of AI-flagged reviews ($N = \nFlagged{}$, score $\geq 5$) against non-flagged low-star reviews ($N = \nNotFlagged{}$) across all \nCategories{} categories. AI-flagged reviews score substantially higher on content quality complaints ($d = \crosstabCQd{}$, $r_{rb} = \crosstabCQrrb{}$, large effect) and presentation complaints ($d = \crosstabPRESd{}$, $r_{rb} = \crosstabPRESrrb{}$). They score substantially lower on meta issues ($d = \crosstabMETAd{}$, $r_{rb} = \crosstabMETArrb{}$), as shown in Figure~\ref{fig:complaint-profile}. AI suspicion co-occurs with content-level quality failures, and with fewer complaints about shipping, titles or formatting.

As a robustness check, we repeated this comparison excluding all Generative AI reviews (removing the \exGenAiFlagsRemoved{} flags from the three most-flagged titles). The directional pattern holds: content quality $d = \exGenAiCQd{}$, presentation $d = \exGenAiPRESd{}$, meta issues $d = \exGenAiMETAd{}$. The presentation and meta effects are larger when Generative AI is excluded, indicating that the co-occurrence pattern is not driven by the high-flag Generative AI titles, and is in fact more pronounced outside them.

AI-flagged reviews also skew toward 1-star ratings: \flaggedOneStarPct{} of flagged reviews are 1-star, compared to \nonflaggedOneStarPct{} of non-flagged low-star reviews.

\subsection{H3 -- Metadata Signals}
\label{sec:rq3}

We tested whether book-level metadata signals predict AI authorship suspicion. Publisher type does. Self-published books show substantially higher mean AI perception scores than books from established publishers ($M_{\text{self}} = \hFourIndieM{}$, $M_{\text{estab}} = \hFourEstabM{}$, $U = \hFourU{}$, $p = \hFourP{}$, $d = \hFourD{}$, large effect) across $N = \hFourIndieN{}$ self-published and $N = \hFourEstabN{}$ established-publisher books. The two review-behavior signals do not: neither 5-star verified purchase rate ($\rho = \hFiveRho{}$, $p = \hFiveP{}$, $N = \hFiveN{}$) nor 5-star review word count ($\rho = \hSevenRho{}$, $p = \hSevenP{}$, $N = \hSevenN{}$) correlates with AI perception. Publisher type is therefore the one book-level signal that carries information, which is consistent with a base-rate account in which self-published titles more often contain AI-generated content.

Publisher type separates ambient from corroborated suspicion. \nFlaggedBooks{} books attract at least one AI-flagged review; of the \flaggedIndieN{} of those whose publisher could be classified, \flaggedIndiePct{} are self-published. Of the \nMultiFlagBooks{} books with three or more flags, \multiFlagIndiePct{} are self-published (two Generative AI titles); the third is a CompTIA certification guide from Packt Publishing, an established technical publisher. The \nEstabFlagged{} established-publisher books that attract AI suspicion (from publishers including Penguin, O'Reilly, Wiley, and Random House) receive at most \estabMaxFlags{} flags each. Corroborated suspicion is concentrated in, but not confined to, self-publishing.

\begin{figure*}[t]
\centering
\includegraphics[width=\textwidth]{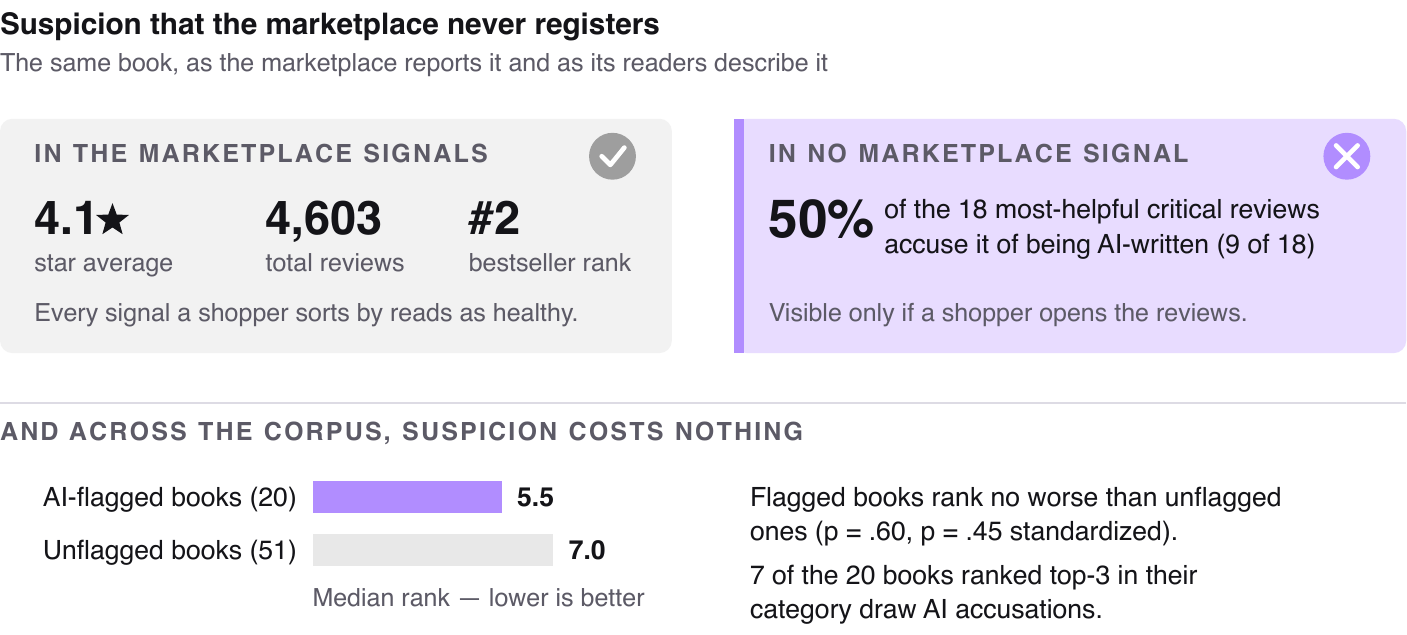}
\caption{Suspicion is visible in the reviews a shopper reads and absent from the signals the marketplace surfaces. Top: the most AI-flagged book in the corpus holds a \mostFlaggedRating{}-star average, \mostFlaggedReviews{} reviews and rank \mostFlaggedRank{} while \mostFlaggedFlagPct{} of its most-helpful critical reviews accuse it of being AI-written. Bottom: across the \hNineFlagN{} flagged and \hNineCleanN{} unflagged books with low-star reviews, flagged books occupy no worse bestseller position ($p = \hNineP{}$ raw, $p = \hNinePPct{}$ standardized within category). The null is conditioned on books that already charted.}
\label{fig:unpunished}
\Description{Two stacked sections. The upper section sets two boxes side by side, separated by the word ``vs''. The left box, headed ``what a shopper sees on the page,'' gives a 4.1 star overall rating, 4,603 total reviews and bestseller rank 2, under the line ``every visible signal reads as a healthy book.'' The right box, headed ``what its critical reviews say,'' gives 50 percent of the most-helpful critical reviews accusing the book of being AI-written, 9 of 18, ranked most helpful by other readers. The lower section is a pair of horizontal bars comparing median bestseller rank, where lower is better: AI-flagged books, 20 of them, at 5.5, and unflagged books, 51 of them, at 7.0. A note records that 7 of the 20 books ranked top three in their category draw AI accusations.}
\end{figure*}

Aggregate and critical-review signals diverge. The most AI-flagged book in the dataset (\textit{The ChatGPT Millionaire}) has \mostFlaggedReviews{} total reviews and a \mostFlaggedRating{}-star overall rating, yet \mostFlaggedFlagPct{} of its most-helpful critical reviews flag AI authorship. These flagged reviews are surfaced by Amazon's ``most helpful'' sort, meaning other readers voted them as useful. The book simultaneously maintains a large, broadly positive readership \textit{and} a visible layer of AI suspicion in its most prominent criticism (Figure~\ref{fig:unpunished}).

This disconnect is not confined to the most-flagged title. Across the \hNineFlagN{} flagged and \hNineCleanN{} unflagged books with low-star reviews, we detect no positional cost to AI suspicion. Flagged books sit at a median bestseller rank of \hNineFlagRank{} against \hNineCleanRank{} for unflagged books ($U = \hNineU{}$, $p = \hNineP{}$), and the comparison remains null after standardizing rank within category to control for uneven sampling depth across categories (\hNineFlagRankPct{} vs.\ \hNineCleanRankPct{}, $U = \hNineUPct{}$, $p = \hNinePPct{}$). \hNineTopThreeFlagged{} of the \hNineTopThree{} books ranked in their category's top three attract AI-flagged reviews. Suspicion is visible in the critical reviews and absent from the position the book occupies. We return to the implications of this disconnect in Section~\ref{sec:discussion}.

\begin{figure*}[t]
\centering
\includegraphics[width=\textwidth]{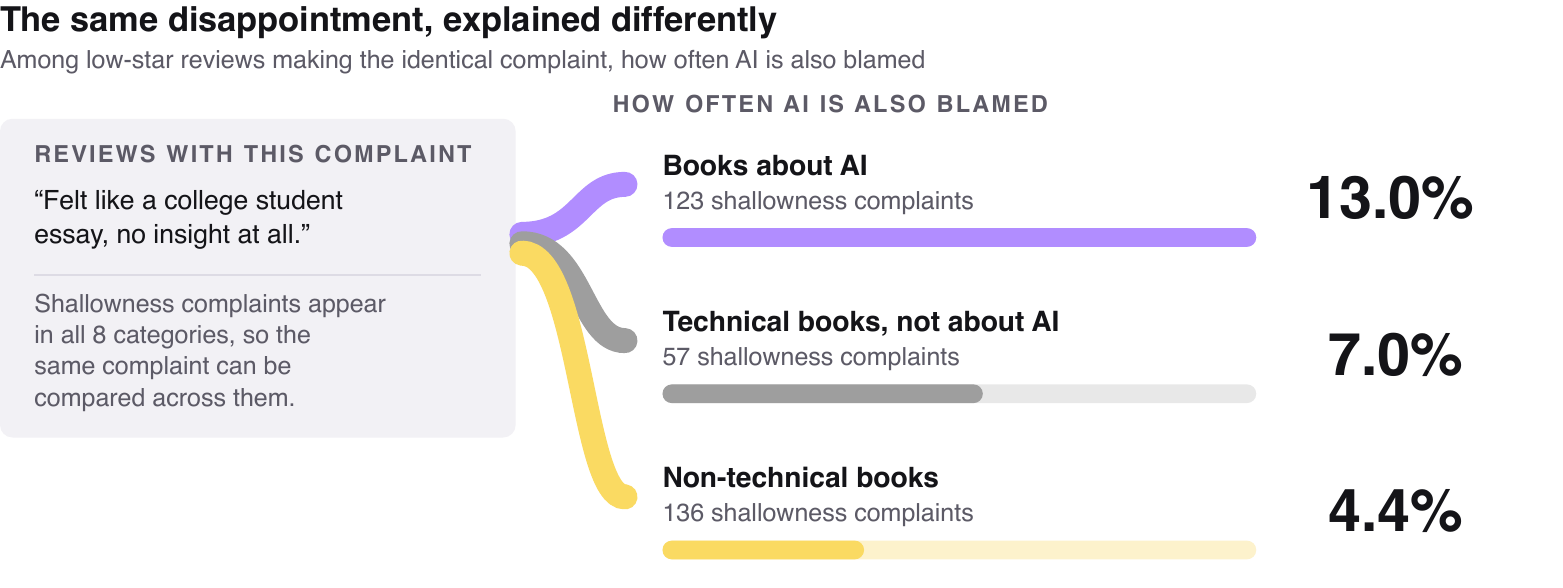}
\caption{The same complaint receives a different explanation depending on the book's subject. Among low-star reviews with a content quality score of 7 or higher (a strong shallowness complaint; quoted: \textit{Co-Intelligence}, AI \& Semantics), \shallowFlagHigh{} also flag AI authorship in high-proximity categories ($n = \shallowNHigh{}$), against \shallowFlagMed{} in medium ($n = \shallowNMed{}$) and \shallowFlagZero{} in zero-proximity categories ($n = \shallowNZero{}$). Holding the complaint fixed removes differing levels of reader disappointment as an explanation; the categories still differ in publisher mix, recency and readership, so this is a conditioned comparison rather than a controlled one.}
\label{fig:translation}
\Description{Two panels joined by three curved connectors. The left panel holds one representative shallowness complaint, quoted as ``Felt like a college student essay, no insight at all,'' with a note that shallowness complaints appear in all eight categories so the same complaint can be compared across them. The connectors run to three horizontal bars on the right, one per proximity tier, showing how often that same complaint also blames AI: books about AI, 123 shallowness complaints, 13.0 percent; technical books not about AI, 57 complaints, 7.0 percent; non-technical books, 136 complaints, 4.4 percent. The bars shorten from top to bottom.}
\end{figure*}

\subsection{Proximity Gradient Analysis}
\label{sec:proximity}

The medium-proximity categories allow a finer-grained test: does AI suspicion track topical proximity to AI, or is technical subject matter sufficient? We group all eight categories into three proximity tiers (high: AI/ML; medium: technical non-AI; zero: non-technical) and test for a monotonic trend in AI perception scores.

The overall trend is significant (Kruskal-Wallis $H = \proxKwH{}$, $p = \proxKwP{}$; Spearman $\rho = \proxRho{}$, $p = \proxRhoP{}$). Mean AI perception by tier: high = \proxHighMean{}, medium = \proxMedMean{}, zero = \proxZeroMean{}. High-proximity categories differ significantly from zero-proximity ($U = \proxHzU{}$, $p = \proxHzP{}$, $d = \proxHzD{}$), but the difference between medium and zero proximity is not significant ($U = \proxMzU{}$, $p = \proxMzP{}$, $d = \proxMzD{}$), and high versus medium falls short of significance ($U = \proxHmU{}$, $p = \proxHmP{}$, $d = \proxHmD{}$). The gradient is driven by AI/ML categories rather than by a smooth increase across tiers. Technical non-AI categories (Computer Security, Programming Languages) show AI perception levels indistinguishable from non-technical categories.

Two cases show that AI suspicion arises outside high-proximity categories, by different routes. Gardening (\flagRateGardening{} flag rate) and Computer Security (\flagRateComputerSecurity{}) show elevated rates despite falling in the zero- and medium-proximity tiers respectively. In Gardening the flags spread across four distinct books: a coloring book with AI-generated images, a gardening handbook by a well-known author suspected of AI ghostwriting, and two guides attributed to apparently unverifiable authors (one self-published, one issued under a single-name imprint). Two of the four, the coloring book and one of the guides, draw the production-evidence heuristics (D6, D3) described in Section~\ref{sec:qualitative}; the handbook and the remaining guide draw unsupported assertion, style deviation, shallow content and poor writing quality.

Computer Security follows a different route. All six flags fall on CompTIA certification prep books, four of them on a single title, and readers cite poor writing quality (D4), prose style (D1), and unsupported assertion (D0) rather than production artifacts; no flagged review in this category invokes D3 or D6. Certification guides are formulaic by design, since their structure follows published exam objectives, and these flags read as quality complaints that reach for AI as the explanation, the entanglement documented in Section~\ref{sec:disc-disappointment} arising without an AI subject matter to prompt it. Artifact detection explains the cross-category baseline in Gardening but not in Computer Security.

\subsubsection{Recency-Matched Comparison.}
The tiers differ in publication recency by construction (Section~\ref{sec:sampling}), with the medium tier the least recent, so we repeat the comparison with all tiers restricted to books published in 2023 or later.

Restricting all tiers to books published in 2023 or later reverses the comparison. Medium-proximity categories flag AI authorship in \recMedFlagRate{} of low-star reviews ($n = \recMedN{}$) against \recZeroFlagRate{} for zero proximity ($n = \recZeroN{}$; $U = \recMzU{}$, $p = \recMzP{}$, $d = \recMzD{}$), and become statistically indistinguishable from high-proximity categories (\recHighFlagRate{}, $n = \recHighN{}$; $p = \recHmP{}$). On a like-for-like comparison, technical subject matter is associated with elevated AI suspicion rather than with the non-technical baseline.

We report both because neither is decisive alone. The full-sample comparison uses all available data but confounds proximity with publication recency; the matched comparison removes that confound but reduces the medium tier to \recMedN{} reviews from eleven books. The unmatched figure is the conservative one, since matching raises the medium tier's rate rather than lowering it, and it leaves the Generative AI concentration untouched: every low-star review in that category comes from a book published in 2023 or later, so its flag rate is identical under both analyses. Together they support a narrower claim than either alone: the Generative AI concentration is robust, while the position of technical non-AI categories relative to non-technical ones is sensitive to recency matching and should be treated as unresolved. Resolving it requires more post-2023 books in those categories than our sample contains.

\section{Discussion}
\label{sec:discussion}

\subsection{A Bimodal Pattern of Reader Suspicion}
\label{sec:disc-bimodal}

Our central finding is that AI authorship suspicion in marketplace reviews follows a bimodal pattern: a dramatic concentration in Generative AI books (\genAiFlagRate{}) coexists with a low but consistent presence (all under 6\%) across every other category we examined, including non-technical ones. Two mechanisms appear to drive these patterns, and they are not mutually exclusive.

\textbf{Topical concentration in AI-related categories.} The most visible part of our data is the concentration in Generative AI. Several non-exclusive factors plausibly contribute: AI/ML categories may genuinely contain more AI-generated content (a base-rate effect), reading about AI may activate ``AI-generated content'' as an applicable category for evaluation~\cite{higgins1977accessibility, tversky1973availability}, and AI/ML readers may have ``AI-generated'' more readily available as critical vocabulary. Our publisher data supports a base-rate component: self-published books attract markedly higher AI perception scores than established-publisher books ($d = \hFourD{}$, $p = \hFourP{}$), the only book-level signal in our data that predicts suspicion. The relationship is not absolute, however: \multiFlagIndiePct{} of the books drawing three or more independent flags are self-published, and the remaining one is a certification guide from an established technical publisher. However, base rates alone do not fully explain the pattern: AI \& Semantics (\flagRateAIAndSemantics{}) and Machine Theory (\flagRateMachineTheory{}) show low flag rates despite likely similar exposure to AI-generated titles. Technical subject matter is a third candidate whose status our data leaves open (Section~\ref{sec:proximity}). Any reading of it in terms of reader expertise is a further inference, since we vary properties of books and not of readers. A specific empirical pattern points toward a topic-amplification component: content shallowness complaints (mean \dimMeanContentQuality{}/10) are widespread across all eight categories, but they translate into an AI flag at different rates by tier. Among low-star reviews with a content quality score of 7 or higher (a strong shallowness complaint), \shallowFlagHigh{} also flag AI authorship in high-proximity categories ($n = \shallowNHigh{}$), against \shallowFlagMed{} in medium ($n = \shallowNMed{}$) and \shallowFlagZero{} in zero-proximity categories ($n = \shallowNZero{}$; $\chi^2(\shallowChiDf{}) = \shallowChi{}$, $p = \shallowChiP{}$; high against zero, $\mathit{OR} = \shallowOR{}$, $p = \shallowORP{}$, Fisher's exact; Figure~\ref{fig:translation}). The same complaint is roughly three times more likely to arrive with an AI explanation attached when the book is about AI. The topic appears to provide interpretive material (an applicable category for explaining quality concerns) that readers of non-AI books do not reach for. One AI/ML reviewer captures this leap directly:

\begin{quote}
\itshape ``A waste of money. Extremely thin, with simple explanations that read like copied from ChatGPT.''\par
\upshape\small\hfill --- 1-star verified review, a machine learning textbook (Machine Theory)
\end{quote}

\noindent The same shallowness complaint in a gardening or history book would rarely arrive with a ChatGPT explanation attached.

\textbf{Direct artifact detection across categories.} The cross-category baseline runs through a different pathway. In Gardening, half the flags follow concrete production evidence (fabricated identities, AI-generated images) on books whose subject matter supplies no AI vocabulary, which is what the D6 and D3 concentration in that category records (Section~\ref{sec:proximity}). Computer Security does not fit this account: its flags cite no production artifact and read instead as quality complaints reaching for AI as an explanation, the entanglement described in Section~\ref{sec:disc-disappointment} operating without an AI subject matter to prompt it. The cross-category baseline is therefore not a single phenomenon.

The two mechanisms likely stack in Generative AI, where books are both about AI and plausibly contain AI-generated content. The Gardening case is informative: it shows that direct artifact detection can produce suspicion without any topical priming, which is difficult to reconcile with a single-mechanism account. Disentangling the relative contributions would require controlled experiments that present identical content with different topic framings.

\subsection{AI Suspicion as an Implicit Disclosure Penalty}
\label{sec:disc-disappointment}

AI suspicion does not occur in isolation. When readers flag AI authorship, their reviews co-express specific quality failures: shallow content ($d = \crosstabCQd{}$), poor presentation ($d = \crosstabPRESd{}$), and \textit{fewer} complaints about misleading titles or formatting ($d = \crosstabMETAd{}$). AI suspicion is entangled with a particular form of disappointment: the reader expected genuine human expertise and received content that feels mass-produced. One review illustrates the full entanglement:

\begin{quote}
\itshape ``Vague and useless. Written by a chatbot? This book doesn't help you understand the subject matter from a technical perspective.''\par
\upshape\small\hfill --- 1-star verified review, a security certification guide (Computer Security)
\end{quote}

\noindent The complaint runs from quality failure (``vague and useless'', ``doesn't help you understand'') to AI authorship as the explanation, with no mention of misleading marketing, formatting, or shipping.

This extends prior work on the AI disclosure penalty~\cite{schilke2025transparency, ragot2020ai, graefe2018perception}, which has shown that \textit{explicit} AI labeling reduces perceived quality. In our data, no labels are present, yet readers reach the same kind of negative judgment via a specific experiential pattern (shallowness, poor writing, lack of grounding). This suggests the disclosure penalty has an implicit counterpart: when content fits a particular failure mode, readers reach for ``AI-generated'' as the explanation, which then carries the same trust penalty that explicit disclosure would impose. The problem readers identify may be content that mimics the surface form of expertise without the substance. A well-edited, AI-assisted book grounded in real experience might not trigger suspicion; a formulaic human-written book might.

\subsection{``AI-Generated'' as Emerging Critical Vocabulary}
\label{sec:disc-vocab}

Ambient suspicion reveals ``AI-generated'' becoming a general-purpose criticism for formulaic writing, regardless of provenance. Its appearance on books that predate ChatGPT (Section~\ref{sec:rq2}) suggests post-hoc reinterpretation rather than detection, though only \preChatGptBookFlags{} reviews provide this evidence directly. The label is sometimes deployed as one of several substitutable explanations for poor writing, as in this review that puts the alternatives side by side:

\begin{quote}
\itshape ``This is either AI slop, poorly edited, or English is not the writer's first language.''\par
\upshape\small\hfill --- 1-star verified review, a security certification guide (Computer Security)
\end{quote}

Our D4 code sits at odds with one prior finding. Russell et al.~\cite{russell2025detectors} asked frequent LLM users to identify AI-generated non-fiction articles; those annotators treated grammatical flawlessness as the mark of machine authorship, describing AI text as ``usually grammatically perfect'' against human writing that ``often contains minor errors'' (24.8\% of their cited cues). Our reviewers make the reverse inference, citing grammar and editing failures as evidence of AI. Why the two diverge is unclear. The settings differ in task, instruction, and text, and the books our reviewers read span several years of model capability, including titles plausibly produced with cheaper or older models whose output is less polished. The grammatical signature readers attribute to AI in this marketplace is not the one that study reports.

\noindent This is consistent with work on folk theories of algorithmic systems~\cite{eslami2016folk}: informal mental models applied broadly, even when inaccurate. The label's emergence as critical vocabulary has structural consequences. Authors in AI-related fields face a Replicant Effect~\cite{jakesch2019replicant}: in mixed environments where AI authorship cannot be verified, suspicion can attach to any individual title without the author's ability to refute it. For human authors writing about AI, the topic itself becomes a liability. The same unverifiability bears on the market as well as the individual title. Books are an information good whose quality buyers cannot assess before purchase~\cite{akerlof1970lemons}, and suspicion is the signal readers generate in place of that assessment. That signal is not priced in: flagged books show no rank disadvantage in our sample (Section~\ref{sec:rq3}), and detected-AI titles take top-rank positions from titles with none~\cite{chakrabarty2026flood}. A marketplace where quality concerns are voiced but carry no positional consequence has no mechanism by which those concerns correct anything.

\subsection{Design Implications}
\label{sec:disc-design}

\textbf{Existing mechanisms are insufficient.} Amazon's KDP AI disclosure policy relies on self-reporting and, at the time of writing, is not surfaced to consumers~\cite{amazon2023kdp, reimers2026ai}. Readers in most categories rarely flag AI authorship, so undisclosed AI content faces little organic accountability outside the highest-visibility cases. The penalty prior work does find attaches to ratings~\cite{reimers2026ai}, leaving the ranking that drives discovery untouched (Section~\ref{sec:disc-vocab}).

\textbf{Aggregate metrics can mask quality concerns.} Current review systems collapse complaints into a single star average. A reader scanning the star average of the most AI-flagged book in our dataset (Section~\ref{sec:rq3}) sees a positive consensus; a reader scanning its ``most helpful'' critical reviews sees a different story. Structured signals that surface this kind of pattern would help readers calibrate before purchase.

\textbf{Aggregation over individual detection.} A single flag carries little information: \dZeroCount{} of the \nHeuristicCoded{} coded reviews assert AI authorship without citing evidence, and \nFlaggedFalsePos{} of the \nFlagged{} flagged reviews are on inspection not contemporary AI authorship claims. Two properties of the corroborated pattern are computable from review text a platform already stores: convergence across independent reviewers, and the class of evidence cited. A book-level signal gated on both would have surfaced two Gardening titles alongside the two most-flagged Generative AI titles, while suppressing isolated assertions. Ungated, the same signal would expose human authors in AI-related categories to the Replicant Effect at platform scale.

\textbf{Transparency signals.} The reader experience our data captures is driven by content quality. Shallowness and lack of grounding draw complaints regardless of how the content was produced, and no available method settles authorship (Section~\ref{sec:limitations}). Platforms can surface disclosure status and structured quality dimensions without making that determination.

\section{Limitations}
\label{sec:limitations}

\textbf{Sample size and category coverage.} Our dataset of \nBooks{} books and \nReviews{} reviews across \nCategories{} categories supports the primary findings but limits generalizability. The medium-proximity tier contains only two categories (\proxMedN{} low-star reviews), and the publisher comparison rests on \hFourIndieN{} self-published books, so both have limited power for pairwise comparisons.

\textbf{Mechanisms cannot be separated.} Two mechanisms operating in parallel is consistent with the bimodal pattern but cannot be isolated by an observational design (Section~\ref{sec:disc-bimodal}). Separating a base-rate account from a priming account would require knowing which books contain AI text, and no available method supplies that. Detectors are a second fallible instrument, so a disagreement between a detector and a reader cannot be attributed to either one, and detector sensitivity varies sharply with how the text was produced: on literary prose, leading detectors flag 97\% of directly prompted output and 3\% of output from a model fine-tuned on an author's works~\cite{chakrabarty2025readers}. Commercially produced AI books plausibly sit closer to the fine-tuned case, so a detector negative would carry little information about whether a reader was mistaken.

\textbf{LLM scorer validity.} Our dimensional scoring uses an LLM (GPT-5.2). We report agreement separately by task type rather than pooling it, because the two are not equally hard. Deciding whether a review asserts AI authorship is an extraction task and agreement is correspondingly high: inter-model ICC = \iccAI{} on that dimension, against \iccOverall{} pooled across all dimensions. Rating how poor a book's content or presentation is, by contrast, is an evaluation task with no textual anchor, and agreement there is much lower for humans and models alike: Shaib et al.~\cite{shaib2025slop} report Cohen's $\kappa$ between $-.15$ and .29 among professional copy-editors judging whether text is ``slop'', and $\kappa = .01$ for GPT-5 on the same judgment. Our high agreement figures are on the extraction task and should not be read as validating the quality dimensions to the same standard. Recent work shows that prompt sensitivity varies across datasets and models, with larger models exhibiting enhanced robustness~\cite{zhuo2024prosa}; we did not formally test prompt sensitivity in our scoring pipeline, but the use of a frontier model with structured JSON output and explicit rubrics is consistent with the conditions under which sensitivity has been observed to be lowest. Egami et al.~\cite{egami2023surrogates} note that even high-accuracy LLM annotation can introduce bias into downstream inference; we report effect sizes alongside $p$-values throughout to support interpretation independent of any single significance threshold.

\textbf{Human validation coverage.} All \humanValFlaggedN{} AI-flagged reviews were annotated by hand, along with \humanValNonFlaggedN{} non-flagged reviews carrying strong complaints on other dimensions. Annotation was performed by an author, blind to LLM scores, dimension rationales, and the review's stratum or category; a multi-annotator design would establish the human-side reliability ceiling more rigorously. Reviews scoring just below the flag threshold are rare in the corpus and are not represented, so agreement is measured away from the decision boundary.

\textbf{Known scorer false positives.} Two concrete failure modes appear in our corpus. First, a review posted in November 2018, four years before ChatGPT's release, scored 7/10 on AI generation perception because it speculated that the author had not personally written the book. In 2018 such speculation concerned human ghostwriting rather than AI. This is the only flagged review posted before ChatGPT's launch (\preChatGptReviewFlags{} of \nFlagged{}). Second, during heuristic coding both coders judged \nHeuristicExcluded{} flagged Motivational Self-Help reviews to express dissatisfaction with writing quality without referencing AI authorship at all. The two modes overlap: the 2018 review is one of the two set aside during coding. Together, \nFlaggedFalsePos{} of \nFlagged{} flagged reviews (\flaggedFalsePosPct{}) are on inspection not contemporary AI authorship claims. This is a more informative false-positive estimate for the flag threshold than the single false positive observed in human validation, and it bounds how much of the low cross-category baseline could be scorer noise rather than reader suspicion.

\textbf{Platform and temporal scope.} We study Amazon US bestseller reviews at a single point in time (February--March 2026). Amazon surfaces some reviews posted on other national marketplaces inline on the US product page, so \nonUsPct{} of our reviews (\nNonUsReviews{}) carry a non-US origin line; we retain them because they are part of what a US shopper reads, but the corpus is therefore not strictly US-authored. Other platforms (Goodreads, Barnes \& Noble) may have different review norms and reader populations. The rapid evolution of AI-generated content means our findings are a snapshot; the prevalence and character of AI suspicion may shift as both AI capabilities and reader awareness change. Longitudinal extension would test whether AI suspicion patterns spread to non-AI categories over time.

\textbf{Bestseller rank is observed only for books that charted.} Our finding that AI-flagged books occupy no worse bestseller positions than unflagged ones (Section~\ref{sec:rq3}) is conditioned on a sample drawn entirely from bestseller lists. Books that failed to chart are not observed, so we can say that suspicion carries no positional cost \textit{among bestsellers}, not that AI-generated content has no market effect. Whether such content expands the market or displaces human-authored titles is a supply-side question requiring sales panel data rather than review text, and our design cannot address it.

\textbf{Review sampling biases.} Our design deliberately oversamples critical reviews and uses Amazon's default sort (most helpful/relevant), which favors recent, articulate reviews. Our flag rates describe the most visible criticism (the reviews most likely to influence prospective buyers), not the overall distribution of reader sentiment. Reviews beyond page 2 of any filter are not captured.

\section{Ethical Considerations}
\label{sec:ethics}

This study analyzes Amazon book reviews posted publicly on product pages. No reviewer was contacted and no account-level data was accessed, and we report no reviewer display names. We quote review text verbatim where a claim rests on the reviewer's own words. Such text is searchable back to its author, so we quote only the passage carrying the claim and identify each quote by the kind of book and its category rather than by title or reviewer.

We name specific book titles when reporting per-book results. This is standard in marketplace research and necessary for reproducibility, but characterizing a smaller or less-established title as attracting AI authorship suspicion could affect its commercial performance more than it would for a high-volume bestseller. Our per-book findings therefore describe what reviewers wrote about those books, and we make no claim about whether any book in our dataset was AI-generated.

We use LLMs to analyze reader suspicion of LLM-generated text. The scoring task, classifying complaint dimensions in review text, is distinct from the object of study, and Section~\ref{sec:limitations} reports what the validation does and does not establish.

\paragraph{Use of generative AI\@.}
Generative AI tools were used in producing this work, which we disclose in accordance with the ACM Policy on Authorship. Claude (Anthropic) was used to build the data collection, scoring, analysis, and figure-generation code, and to draft sections of the text, which the authors revised. Research questions, study design, hypotheses, the heuristic codebook, human annotation, and all editorial decisions are the authors' own, and the authors take responsibility for all content.

No IRB review was required for this study, as it analyzes publicly available text data with no human subjects interaction. The study was conducted in accordance with the ACM Code of Ethics.

\section{Conclusion}
\label{sec:conclusion}

We studied how AI authorship suspicion manifests in book marketplace reviews across \nReviews{} Amazon reviews of \nBooks{} bestselling books in \nCategories{} categories at three levels of topical proximity to AI. AI authorship suspicion follows a bimodal pattern: a dramatic concentration in Generative AI books (\genAiFlagRate{} of critical reviews flag AI authorship) coexists with a low but consistent presence (all under 6\%) across every other category, including non-technical ones. The Gardening category, with zero topical proximity to AI, shows the second-highest flag rate (\flagRateGardening{}), driven by concrete artifacts: phantom authors, AI-generated imagery, and suspected ghostwriting in established titles. Across all books, Computer Security and Programming Languages sit within the same range as non-technical categories, though this contrast reverses when tiers are matched on publication recency, leaving the role of technical subject matter unresolved.

Two mechanisms appear to drive these patterns: topical concentration, where AI subject matter raises both the plausible share of AI-generated titles and readers' readiness to name AI as the cause of quality complaints, and direct artifact detection, which surfaces across categories regardless of topic. Our observational design surfaces these patterns but cannot experimentally isolate the mechanisms. When readers do suspect AI authorship, their reviews complain more about shallow content ($d = \crosstabCQd{}$) and poor presentation ($d = \crosstabPRESd{}$), and less about misleading titles or formatting ($d = \crosstabMETAd{}$). Suspicion takes two qualitatively distinct forms: \textit{ambient}, where ``AI-generated'' serves as generic criticism regardless of provenance, and \textit{corroborated}, where multiple independent reviewers flag the same book with specific evidence. Publisher type is the one book-level signal that predicts suspicion (self-published $d = \hFourD{}$, $p = \hFourP{}$), though corroborated suspicion appears in established-publisher titles as well.

Outside Generative AI, reader reviews rarely flag AI authorship, and aggregate metrics can mask the suspicion that does appear in the most-helpful critical reviews. Marketplace design should move toward structured quality signals, transparent AI disclosure, and aggregated suspicion metrics rather than relying on organic detection alone. Conducted approximately three years after the launch of ChatGPT, this study contributes a dated empirical baseline against which future work can track how reader responses to AI-generated content evolve as AI capabilities, content volume, and reader awareness change.

\paragraph{Data and Code Availability.}
The dataset of \nReviews{} dimensionally scored reviews of \nBooks{} bestselling books will be released under CC BY-NC 4.0; analysis code, scoring prompts, and the replication protocol under MIT. Held-out train/dev/test splits, official benchmark tasks, evaluation metrics, and a baseline distilled scorer will be released alongside it.

%% References
\bibliographystyle{ACM-Reference-Format}
\bibliography{references}

%% Appendix
\clearpage
\appendix
\section{Category Selection Protocol}
\label{app:category-selection}

Table~\ref{tab:category-discovery} shows the candidate categories evaluated during zero-proximity category selection. These ratios were recorded at discovery time from bestseller pages that change daily, and were used only to guide category selection; they are not recomputed from the retained corpus and do not enter any analysis. Publisher composition of the books actually collected is reported in Section~\ref{sec:rq3}.

\begin{table}[h]
\caption{Category discovery results for zero-proximity category selection. Established-to-indie ratio and recent book percentage were computed from the top 30 bestsellers in each subcategory.}
\label{tab:category-discovery}
\resizebox{\narrowwidth}{!}{%
\begin{tabular}{lrrl}
\toprule
Category & Est:Ind & Recent \% & Selected \\
\midrule
Bus. Dev. \& Entrepreneurship & 2:1 & $\sim$40\% & No\textsuperscript{a} \\
Personal Finance & 9:1 & $\sim$20\% & No\textsuperscript{b} \\
American History & 14:1 & $\sim$50\% & Axis 2 \\
Motivational Self-Help & 3:1 & $\sim$35\% & Axis 1 \\
\bottomrule
\end{tabular}}
\begin{flushleft}
\small
\textsuperscript{a}Risk of AI subject-matter overlap (``use ChatGPT for business''). \\
\textsuperscript{b}Dominated by pre-2023 classics; insufficient recent titles.
\end{flushleft}
\end{table}

\section{Scoring Prompt}
\label{app:scoring-prompt}

The full scoring prompt used for low-star review dimensional scoring:

\begin{quote}
\small\sloppy
Score this Amazon book review on each dimension below. For each dimension, assign an integer score from 1 to 10: 1 = completely absent from the review, 3 = weak/slight signal, 5 = moderate signal, 7 = strong signal, 10 = dominant theme of the review. Also provide a one-sentence rationale for each score.

DIMENSIONS:
\begin{itemize}
    \item \textbf{content\_quality}: Reviewer thinks the content is shallow, surface-level, lacks depth, is unoriginal/rehashed from free sources, or is artificially padded.
    \item \textbf{accuracy}: Reviewer flags factual errors, code that doesn't work, incorrect technical claims, misleading information.
    \item \textbf{ai\_generation\_perception}: Reviewer explicitly suspects or states that the BOOK CONTENT was generated by AI, ChatGPT, or an LLM. IMPORTANT: Only score high if the reviewer suspects AI-generated authorship. Do NOT score high just because the book's SUBJECT MATTER is about AI/ChatGPT/LLMs.
    \item \textbf{presentation}: Reviewer criticizes writing quality, grammar, typos, repetition, poor organization, or outdated content.
    \item \textbf{meta\_issues}: Non-content problems: misleading title/marketing, wrong audience fit, shipping damage, Kindle formatting, physical defects.
\end{itemize}
\end{quote}

\section{Per-Book AI Perception Scores}
\label{app:per-book}

Table~\ref{tab:per-book} shows the top 20 books ranked by mean AI generation perception score.

\begin{table}[h]
\caption{Top 20 books by mean AI generation perception score (low-star reviews). Flag rate = proportion of reviews scoring $\geq 5$.}
\label{tab:per-book}
\resizebox{\narrowwidth}{!}{%
\begin{tabular}{lllrrr}
\toprule
Title & Category & $N$ & Mean & Flag \% \\
\midrule
The ChatGPT Millionaire: \ldots & Generative AI & 18 & 4.83 & 50 \\
Build Passive Income with\ldots & Generative AI & 11 & 3.36 & 27 \\
THE AI WORKSHOP: Your Com\ldots & Generative AI & 4 & 3.25 & 25 \\
CompTIA® Security+® SY0-7\ldots & Computer Security & 18 & 2.89 & 22 \\
CompTIA Security+ SY0-701\ldots & Computer Security & 5 & 2.80 & 20 \\
Former Amish Reveals: The\ldots & Gardening & 16 & 2.19 & 12 \\
Martha Stewart''s Gardeni\ldots & Gardening & 19 & 1.95 & 11 \\
Tiny Gardens: Cute \& Comf\ldots & Gardening & 19 & 1.95 & 11 \\
Herbal Synergies: A Guide\ldots & Gardening & 18 & 1.83 & 11 \\
The Coming Wave: AI, Powe\ldots & AI \& Semantics & 20 & 1.75 & 10 \\
50 Algorithms Every Progr\ldots & Prog. Languages & 9 & 1.67 & 11 \\
Always Remember: The Boy,\ldots & Motiv. Self-Help & 14 & 1.57 & 7 \\
Advances in Financial Mac\ldots & Machine Theory & 18 & 1.50 & 6 \\
The Hundred-Page Machine \ldots & Machine Theory & 18 & 1.50 & 6 \\
1929: Inside the Greatest\ldots & Amer. History & 19 & 1.47 & 5 \\
AI Engineering: Building \ldots & Machine Theory & 17 & 1.47 & 6 \\
If Anyone Builds It, Ever\ldots & AI \& Semantics & 15 & 1.47 & 7 \\
CompTIA Security+ Get Cer\ldots & Computer Security & 15 & 1.47 & 7 \\
Co-Intelligence: Living a\ldots & AI \& Semantics & 18 & 1.44 & 6 \\
The Laws of Human Nature & Motiv. Self-Help & 17 & 1.35 & 6 \\
\bottomrule
\end{tabular}}
\end{table}

Of the 78 books, 51 have mean AI perception scores below 1.3 (near floor). The top three books (all in Generative AI and independently published) account for 13 of the 38 AI-flagged reviews in the full dataset (34\%).

\section{Inter-Model Validation Details}
\label{app:validation}

Table~\ref{tab:intermodel} reports per-dimension agreement between the primary scorer (GPT-5.2) and the validation scorer (Grok-4-1) on a stratified 20\% sample ($N = 379$ reviews, 1,753 score pairs).

% Auto-generated by make_intermodel_table.py. Do not edit manually.
\begin{table}[h]
\caption{Inter-model agreement by dimension. ICC = intraclass correlation ICC(2,1); $\kappa_w$ = weighted Cohen's kappa; MAD = mean absolute difference.}
\label{tab:intermodel}
\resizebox{\narrowwidth}{!}{%
\begin{tabular}{lrrrrr}
\toprule
Dimension & $N$ & ICC & $\kappa_w$ & MAD & Exact \% \\
\midrule
AI perception & 237 & 0.969 & 0.939 & 0.03 & 99.2 \\
Accuracy & 237 & 0.915 & 0.847 & 0.41 & 78.1 \\
Presentation & 379 & 0.874 & 0.782 & 0.93 & 52.5 \\
Content quality & 379 & 0.870 & 0.733 & 1.08 & 43.3 \\
Practical value & 142 & 0.859 & 0.758 & 1.24 & 35.2 \\
Meta issues & 237 & 0.846 & 0.697 & 1.24 & 50.2 \\
Logistical only & 142 & 0.683 & 0.558 & 0.23 & 95.1 \\
\midrule
\textbf{Overall} & \textbf{1{,}753} & \textbf{0.902} & \textbf{0.804} & \textbf{0.78} & \textbf{62.0} \\
\bottomrule
\end{tabular}}
\end{table}

The AI generation perception dimension shows the strongest agreement of any dimension (ICC = \iccAI{}, exact agreement = \imExactAI{}). This is partly because AI suspicion is rare (most reviews score 1), but the models also agree on the few reviews that score high: in the Generative AI category, where variance is highest, ICC = \imIccGenAI{} with \imWithinOneGenAI{} within-1 agreement. Systematic bias is small: Grok-4-1 scores average \imBias{} points higher than GPT-5.2 across dimensions, consistent across categories. This level of bias does not affect ordinal comparisons or threshold-based classification.

\section{Qualitative Coding Inter-Rater Agreement}
\label{app:qualcoding}

The qualitative heuristic coding (Section~\ref{sec:qualitative}) was performed independently by two models on \nHeuristicCoded{} of the \nFlagged{} AI-flagged reviews. Two flagged reviews from Motivational Self-Help were excluded because they expressed dissatisfaction with writing quality without referencing AI authorship. Table~\ref{tab:qualagree} summarizes agreement.

\begin{table}[h]
\caption{Inter-rater agreement for qualitative heuristic coding ($N = \nHeuristicCoded{}$).}
\label{tab:qualagree}
\begin{tabular*}{\narrowwidth}{@{\extracolsep{\fill}}lr}
\toprule
Metric & Value \\
\midrule
Cohen's $\kappa$ & \heuristicKappa{} \\
Weighted $\kappa$ & \heuristicWeightedKappa{} \\
Exact agreement & \heuristicExactPct{} (\heuristicExactN{}/\nHeuristicCoded{}) \\
Disagreements & 6/\nHeuristicCoded{} \\
\bottomrule
\end{tabular*}
\end{table}

Four of the six disagreements involved adjacent evidence-based heuristics (e.g., D1/prose style vs.\ D2/shallow content) rather than fundamentally different interpretations. The remaining two concerned whether a terse accusation counted as unsupported (D0) or as citing evidence: one review (``AI Generated trash'') was coded D0 by one model and D4/poor writing by the other, and another (``This book may have been created by AI\ldots'') was coded D5/factual errors against D0. The boundary between unsupported and minimally evidenced accusations is therefore not perfectly sharp, which bears on the \dZeroSharePct{} unsupported-assertion share reported in Section~\ref{sec:qualitative}.

\section{Language of AI Suspicion}
\label{app:phrases}

Table~\ref{tab:phrases} shows the key phrases reviewers use when expressing AI authorship suspicion. Terms fall into two groups: \textit{AI-attribution} terms (how reviewers name the suspicion) and \textit{quality-complaint} terms (what they co-express alongside AI suspicion). The generic term ``AI'' appears most frequently, followed by specific model references (``ChatGPT,'' ``GPT''). Quality complaints co-occurring with AI suspicion focus on wasted value (``waste of money/time'') and content emptiness (``shallow,'' ``fluff'').

\begin{table}[h]
\caption{Key phrases in AI-flagged reviews ($N = 38$). AI-attribution terms (how reviewers name the suspicion) and quality-complaint terms (what they co-express). Counts reflect mentions across review titles and body text.}
\label{tab:phrases}
\resizebox{\narrowwidth}{!}{%
\begin{tabular}{llrr}
\toprule
Phrase & Type & Count & \% of reviews \\
\midrule
AI & AI & 66 & 174\% \\
ChatGPT / GPT & AI & 35 & 92\% \\
AI-generated / AI generated & AI & 12 & 32\% \\
written by AI & AI & 5 & 13\% \\
artificial intelligence & AI & 3 & 8\% \\
AI slop & AI & 2 & 5\% \\
waste (of money/time) & Quality complaints & 7 & 18\% \\
shallow / generic / vague & Quality complaints & 6 & 16\% \\
fluff / filler / slop & Quality complaints & 7 & 18\% \\
poorly written & Quality complaints & 1 & 3\% \\
garbage / trash / useless & Quality complaints & 4 & 11\% \\
doesn't exist (author) & Quality complaints & 1 & 3\% \\
\bottomrule
\end{tabular}}
\end{table}

\end{document}